\documentclass[final,authoryear,5p,times,twocolumn]{elsarticle}
\usepackage{color}
\usepackage{graphics}
\usepackage{txfonts}
\usepackage[breaklinks=true,draft]{hyperref}
\usepackage{multirow}

\journal{Astronomy and Computing}

\newcommand{\urlsamefont}[1]{\urlstyle{same}\url{#1}}

\begin{document}

\begin{frontmatter}

\title{The Virtual Astronomical Observatory:  Re-engineering Access to Astronomical Data}

\author[VAO]{Hanisch, R.~J.\fnref{a}}
\fntext[a]{Current address:  Office of Data \& Informatics, Material Measurement Laboratory, National Institute of Standards \& Technology, Gaithersburg, MD  20899 USA}
\ead{hanisch@stsci.edu}

\author[IPAC]{Berriman, G.~B.}
\ead{gbb@ipac.caltech.edu}

\author[JPL]{Lazio, T.~J.~W.}
\ead{joseph.lazio@jpl.nasa.gov}

\author[CACR]{Emery Bunn, S.}
\ead{sarah.emery@caltech.edu}

\author[SAO]{Evans, J.}
\ead{janet@cfa.harvard.edu}

\author[GSFC]{McGlynn, T.~A.}
\ead{thomas.a.mcglynn@nasa.gov}

\author[UIUC]{Plante, R.}
\ead{rplante@illinois.edu}

\address[VAO]{Virtual Astronomical Observatory, 1600 $14^{\mathrm{th}}$ Street \hbox{NW}, Suite~730, Washington, DC 20036  USA and\\ Space Telescope Science Institute,
3700 San Martin Drive, Baltimore, MD 21218 USA}

\address[IPAC]{Infrared Processing and Analysis Center, California
Institute of Technology, Pasadena, CA 91125  USA}

\address[JPL]{Jet Propulsion Laboratory, California Institute of
Technology, Pasadena, CA 91109 USA}

\address[CACR]{Center for Advanced Computing Research, California
Institute of Technology, Pasadena, CA 91125}

\address[SAO]{Smithsonian Astrophysical Observatory, 60 Garden Street,
Cambridge, MA 02138 USA}

\address[GSFC]{High Energy Astrophysics Science Archive Research Center,
NASA Goddard Space Flight Center, Greenbelt, MD 20771 USA}

\address[UIUC]{National Center for Supercomputing Applications, University
of Illinois, Urbana, IL 61801  USA}

\begin{abstract}
The U.{}S.\ Virtual Astronomical Observatory was a software infrastructure and development project designed both to begin the establishment of an operational Virtual Observatory (VO) and to provide the U.{}S.\ coordination with the international VO effort.
The concept of the VO is to provide the means by which an astronomer is able to discover, access, and process data seamlessly, regardless of its physical location.  This paper describes the origins of the VAO, including the predecessor efforts within the U.{}S.\ National Virtual Observatory, and summarizes its main accomplishments.  These accomplishments include the development of both scripting toolkits that allow scientists to incorporate VO data directly into their reduction and analysis environments and high-level science applications for data discovery, integration, analysis, and catalog cross-comparison.  Working with the international community, and based on the experience from the software development, the VAO was a major contributor to international standards within the International Virtual Observatory Alliance.  The VAO also demonstrated how an operational virtual observatory could be deployed, providing a robust operational environment in which VO services worldwide were routinely checked for aliveness and compliance with international standards.  Finally, the VAO engaged in community outreach, developing a comprehensive web site with on-line tutorials, announcements, links to both U.{}S.\ and internationally developed tools and services, and exhibits and hands-on training at annual meetings of the American Astronomical Society and through summer schools and community days.  All digital products of the VAO Project, including software, documentation, and tutorials, are stored in a repository for community access.  The enduring legacy of the VAO is an increasing expectation that new telescopes and facilities incorporate VO capabilities during the design of their data management systems.
\end{abstract}


\end{frontmatter}

\section{Introduction}\label{sec:introduction}

\subsection{Beginnings}
The formal Virtual Observatory (VO) program in the United States began
with the 2000 Decadal Survey of the National Academy of Science, in
which a National Virtual Observatory (NVO) was identified as the top
priority small initiative \citep{AANM}.

\begin{quote} \em
The NVO is the committee's top-priority small initiative. NVO involves the integration of all major astronomical data archives into a digital database stored on a network of computers, the provision of advanced data exploration services for the astronomical community, and the development of data standards and tools for data mining. É The committee recommends coordinated support from both NASA and the \hbox{NSF}, since NVO will serve both the space- and ground-based science communities.
\end{quote}

The NVO project and parallel projects in Europe and the U.{}K.\ were
formulated through a series of meetings, beginning with ``Virtual
Observatories of the Future'' \citep{2001ASPC..225.....B}, held at the
California Institute of Technology in~2000 June.

At the 2002 conference, ``Toward an International
Virtual Observatory'' \citep{2004tivo.conf.....Q}, held in Garching,
Germany, the International Virtual Observatory Alliance\footnote{
\texttt{http://www.ivoa.net/}}
(IVOA) was formed with the \hbox{NVO}, {the Astrophysical
Virtual Observatory (\hbox{AVO}, ESO)}, and AstroGrid (U.{}K.) as
founding partners. R.~Hanisch, the then-NVO Project Manager, was the
first chair of the IVOA Executive Committee.  In the subsequent
decade, the IVOA has grown to have 21 member national projects.

The IVOA patterned itself on the World-Wide Web Consortium\footnote{
\texttt{http://www.w3.org/}
}
(W3C) and adopted its process for the development of standards
(Working Drafts $\rightarrow$ Proposed Recommendations $\rightarrow$
Recommendations) with the actual standards documents developed by a
set of working groups.  (See \S\ref{sec:voarch} for more details.)
A Virtual Observatory Working Group was established
under Commission~5 of the International Astronomical Union (IAU) in
order to give IVOA Recommendations official status within the
\hbox{IAU}, but this process {has not been used in practice} since there was already global acceptance of IVOA standards.

The NVO project focused on standards and infrastructure development,
working closely in the context of the \hbox{IVOA}, and implemented a
number of prototype science applications to demonstrate the utility of
the underlying VO standards. NVO also ran an active program of
engagement with the astronomical community through annual summer
schools of one-week duration, exhibits at American Astronomical
Society meetings, and the production of a major reference
book, \textit{The National Virtual Observatory:  Tools and Techniques for
Astronomical Research} \citep{2007ASPC..382.....G}.  In a
demonstration of this book's value, it was translated into Mandarin by members of the VO-China project.  

The NVO project was funded by
the National Science Foundation's Information Technology Research
program, starting in~2001, and included organizations in astronomy and
computer science.  Its 
funding came to a planned close in~2008, after demonstrating the technology framework for supporting a \hbox{VO}.

\subsection{Program}

In~2010, the successor to the \hbox{NVO}, the Virtual Astronomical
Observatory (VAO), was begun to sustain and evolve those technologies
successfully demonstrated by the NVO as part of an operating virtual
observatory.  While there were numerous management and logistical
barriers to the establishment of the \hbox{VAO}, the {National
Science Foundation} (NSF) and {National Aeronautics \& Space
Administration} (NASA)
agreed to fund the project jointly, with NSF support directed through
the 
{\hbox{VAO}, Limited Liability Company}, 
and NASA support provided directly to the participating NASA data centers.

The \hbox{VAO}, \hbox{LLC}, was created as a 50-50 collaboration
between the Association of Universities for Research in Astronomy
(AURA) and the Associated Universities, Inc.\ (AUI), with an
independent Board of Directors. This management structure was chosen
deliberately so that the VAO would be perceived as belonging to the
research community and have dedicated oversight.  
Executive authority within the VAO was provided by the Director, who
worked with a Program Manager, Project Scientist, and Project Technologist.
In order to provide
advice on priorities for research tools, a Science Council was
established.  Within the \hbox{VAO}, a Program Council consisting of
senior management representatives from each VAO member organization
was also established.
The Program Council worked with the VAO
management  to map Science Council priorities onto available resources and expertise, and thus to develop the annual program plan. Work packages for all organizations, whether funded by NSF or \hbox{NASA}, were agreed with the Director and Program Manager. The program plan covered all work at all organizations regardless of the source of funding.

\begin{table*}[t]
\centering
\caption{VAO Funding and Review History\label{tab:funding}}
\begin{tabular}{lcl}
\noalign{\hrule}
2010 Apr & NSF Cooperative Agreement issued & \$2M NSF + \$1.5M NASA (FY10) \\
         &                                  & \$4M NSF + \$1.5M NASA (FY11) \\
2010 Aug & PEP v1.0       & \\
2010 Oct & PEP v1.1       & \\
2011 Apr & PEP and review                   & \$2M NSF + \$1M NASA (FY12) \\
2012 Feb & PEP v2.0       & \\
2012 Mar & PEP v2.1, v2.2 & \\
2012 May & PEP v2.3       & \\
2012 Jul & PEP and review & \\
2012 Sep & decision to terminate \hbox{VAO}, & \$2M NSF + \$1M NASA (FY13) \\
  &  effective 2014 September                & \$1M NSF + \$0.5M NASA (FY14) \\
  & \textbf{Total Funding}                       & \textbf{\$11M NSF + \$5.5M NASA} \\
\noalign{\hrule}
\end{tabular}
\\
\parbox{0.7\textwidth}{PEP refers to the Project Execution Plan, an annual deliverable to the funding agencies.  NSF's funding vehicle was a Cooperative Agreement (CA) with the \hbox{VAO}, \hbox{LLC}.}
\end{table*}

Table~\ref{tab:funding} shows the VAO program history and funding. As
a result of two major reviews, NSF and NASA redefined program
priorities and reduced the overall budget from an original plan of
\$27.5M (\$20M NSF + \$7.5M NASA) to~\$16.5M (\$11M NSF + \$5.5M
NASA).  {In addition to simple reductions in funding, these reviews
were often accompanied by recommended changes in the direction of the
project, and, ultimately, the project duration was reduced by seven
months.  Consequently, some activities that were started or intended
to be started were reduced in scope or stopped early to respond to the
combination of lower funding and recommended changes in direction.  A
specific example of this change in direction and cessation of
activities was the Time Series Search Tool (\S\ref{sec:time}), which
was unable to be brought to the desired level of maturity.}

\subsection{Major Accomplishments}
The accomplishments of the NVO and VAO are extensive and will be described in further detail in the following sections of this paper. At a summary level, however, we note the following accomplishments:

\begin{itemize}
\item Major contributor to IVOA standards.  \ref{app:standards} contains a list of IVOA standards to which NVO/VAO staff contributed. The list includes standards recommended by the IVOA Executive Committee and those submitted to the Executive Committee for recommendation. 
\item Leadership within the \hbox{IVOA}, within the executive, Working Groups, and Interest Groups.
\item High-level science applications for data discovery, integration, analysis, and catalog cross-comparison.
\item Scripting toolkits that allow scientists to incorporate VO data directly into their reduction and analysis environments.
\item A robust operational environment in which VO services worldwide are routinely checked for aliveness and compliance with IVOA standards.
\item Community engagement through AAS meetings, summer schools (NVO), and community days (VAO).
\item Comprehensive web site with on-line tutorials, announcements, links to both U.{}S.\ and internationally developed tools and services.
\item Take up of VO standards and infrastructure within essentially every major data center and survey project in the United States, with approximately 1M VO-based data requests per month 
and some 2000 unique users.

\item Prudent fiscal management, with overall management expenses kept
below~15\% and the project completed with an unspent balance of funds
of less than 1\% (for an \$11M [lifetime] budget over~4~years).
\end{itemize}

\section{Science Applications}\label{sec:SciApps}

The VAO developed three science applications (Data Discovery Tool,
Iris Interoperable SED Access and Analysis tool, and the Catalog Cross
Comparison Service) and one prototype application (Time Series Search
tool), {all described in more detail below}.  {There
were also various community-led efforts, that while not formal VAO
projects, built upon VO standards and often involved VAO personnel in
other capacities.  These are also summarized below.}

{The motivations for developing these science applications were
two-fold.  First, before a standard is adopted as an IVOA
Recommendation, it is expected that the Working Draft have \emph{two}
reference implementations.  The objective is to ensure that the
intentions of standards actually can be met in practice.  In developing
these science applications, the VAO provided feedback to the larger
IVOA community on various aspects of IVOA standards.  Second, these
science applications were developed in concert with the research
community, providing additional or new capabilities for addressing a
variety of astronomical research questions.}
{In the spirit that the VO is intended to enable data discovery
and access for all astronomers, the} applications do not serve any one observatory, wavelength, or type of user, but were intended for use by astronomers with multi-wavelength data from possibly a variety of telescopes that span the electromagnetic spectrum.

{As part of a larger goal of developing an environment or
``ecosystem'' in which astronomical software can interact seamlessly
and other tools can be contributed by the community, the development
path for these science applications often included making them
interoperable with other VO tools.  In so doing, the VAO also provided
feedback to the IVOA on the approaches toward interoperability.
As a consequence of developing these applications, a number of
libraries or services were developed that enable other developers to
add functionality to the applications. Two examples are the SEDLIB
(SED I/O library) and NED/SED service developed for Iris.  Finally, 
by way of encouraging contributions, several collaborations (e.g., ASI Science Data Center (ASDC) archive plug-in for Iris) were fostered during VAO science applications development.
}

\subsection{Data Discovery Tool}
The Data Discovery Tool (DDT) is a web application for discovering all
resources about an astrophysical object or a region
of the sky (\S\ref{sec:voarch}).  Using protocols defined by the \hbox{IVOA}, the DDT
searches those widely distributed resources that are found in the VO Registry and presents the results
in a single unified Web page.  In the spirit of the VAO being a
working astrophysical observatory, the DDT was designed to serve as
the initial steps toward a ``portal,'' a means of discovering and
accessing multi-wavelength data.

\begin{figure}[tb]
\centering
\resizebox{\hsize}{!}{\includegraphics{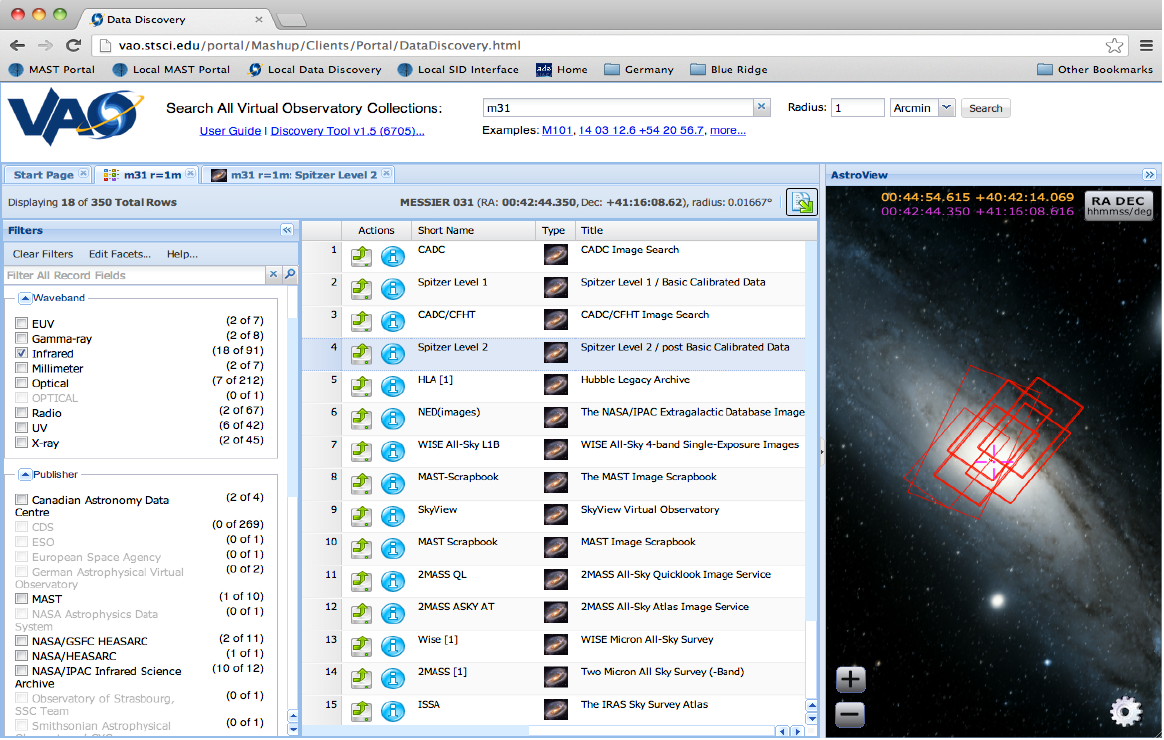}} 
\caption{Appearance of the Data Discovery Tool (DDT) after a search for M31 with a radius of~$1^\prime$ showing the filters (left panel) that can be applied to the search results (center panel), and the AstroView component with field-of-view overlays representing the available data sets.}
\label{fig:DDT}
\end{figure}

Many of the most popular U.{}S.\ archives and catalog holdings are
available for searches in the \hbox{DDT}, including the \textit{Hubble
Space Telescope}, \textit{Chandra} X-ray Observatory, the Mikulski
Archive for Space Telescopes (MAST), the High Energy Astrophysics
Science Archive Research Center (HEASARC), Sloan Digital Sky Survey
({SDSS}), \textit{Spitzer} Space Telescope, and the Two Micron All Sky
Survey ({2MASS}), to name a few. A powerful filtering mechanism allows the user to quickly narrow the initial results to a short list of likely applicable data. Guidance on choosing appropriate data sets is provided by a variety of integrated displays, including an interactive data table, basic histogram and scatter plots, and an all-sky browser/visualizer with observation and catalog overlays (Figure~\ref{fig:DDT}).

The DDT was developed incrementally with the first release of the
application in~2011 June. Development continued over the next two years with five incremental releases that added features and addressed any deficiencies.  Web-based user documentation and training videos were developed and updated for each release.

The DDT project utilized DataScope \citep[\hbox{GSFC/NVO},
][]{2007ASPC..382...51M} and Astroview (STScI) and shared synergy with
the MAST archive development project at \hbox{STScI}. IVOA standards
feedback was substantial.
Experience from the DDT project was used to advocate for
enhanced registry metadata, table access protocol improvements, and enhanced data
access protocols to ensure support for bulk queries.  Staff involved in DDT 
development also helped to write the IVOA standard on HEALPix Multi-Order Coverage maps \citep{IVOA_MOC} for describing sky coverage.

\subsection{Interoperable SED Access and Analysis Tool, Iris}

\begin{figure}[tb]
\centering
\resizebox{\hsize}{!}{\includegraphics{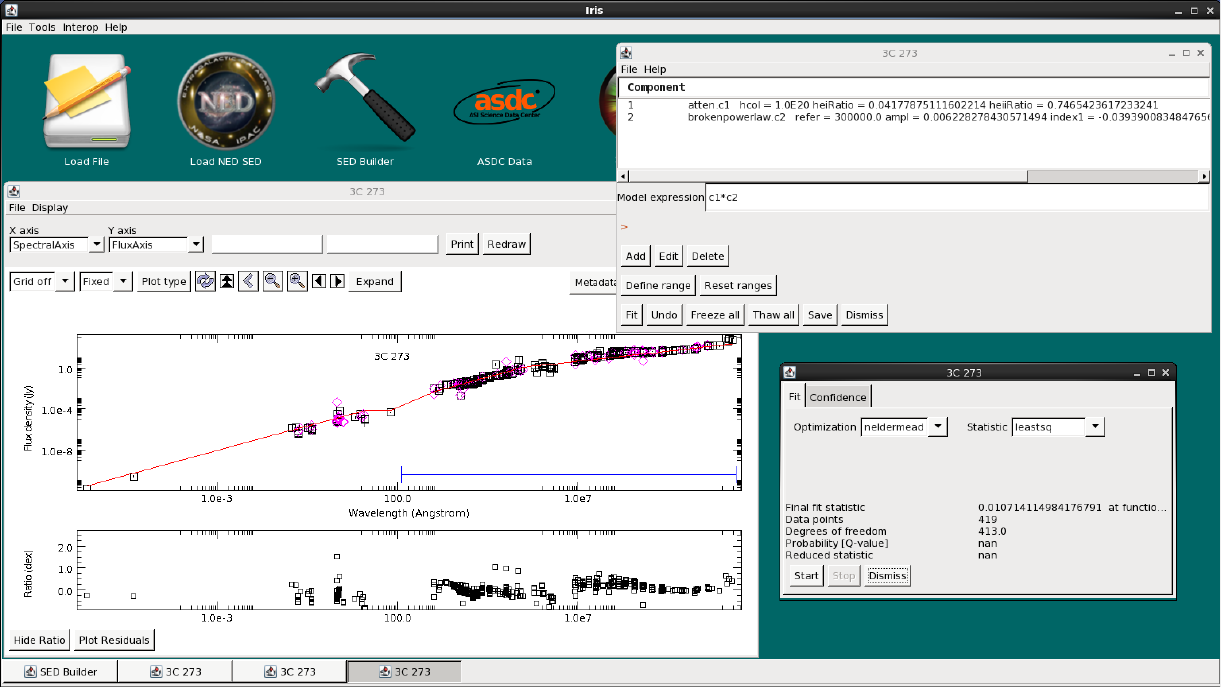}} 
\caption{VAO SED access and analysis tool Iris in operation.  The Iris
desktop holds the interactive windows for SED data review and
analysis.  Shown is a panel displaying the SED of 3C~273 with a model
fit (red curve) and two panels from which the user can describe the
model to fit an SED and control the fitting.}
\label{fig:Iris}
\end{figure}

Iris is a downloadable Graphical User Interface application that
enables astronomers to build and analyze wide-band spectral energy
distributions
\citep[SEDs,][]{2012ASPC..461..893D,2014ASPC..485...19L,2014A&C.....7...81L}.
SED data may be loaded into Iris from a file on
the user's local disk, from a remote \hbox{URL}, or directly from the NASA
Extragalactic Database (NED) for analysis via the NED/SED Service. A
plug-in component enables users to extend the functions of
Iris. Iris utilized Sherpa
\citep{2001SPIE.4477...76F,2006ASPC..351...77D} and Specview \citep{2002SPIE.4847..410B} as
the components that performed fitting and visualization in the
application. Communication between Specview and Sherpa is managed by a
Simple Application Messaging Protocol (SAMP) connection
\citep{IVOA_SAMP,2012ASPC..461..279T}.  Data can also
be read into Iris and can be written out via the SAMP interface
\citep{2012ASPC..461..391L}.  
A separable library for SED data input/output (SEDLib) is also included and available independently from Iris (Figure~\ref{fig:Iris}).

Iris was first released in~2011 October. Three incremental releases
and one bug fix release followed.  Iris is supported on several
versions of the Mac OS~X and Linux. Web-based documentation and
user training videos are also provided. Iris was featured
on the Astrobetter blog in~2013 September.\footnote{
\texttt{http://www.astrobetter.com/\\ release-iris-2-0-sed-analysis-tool/}}

There were two by-products of the Iris project---the NED/SED service
and the SEDLib. There were collaborations with several groups
including the ASI Science Data Center (ASDC) and CDS (Strasbourg). The
collaborations led to Iris desktop plug-in services to access the
respective SED data holdings \citep{2013ASPC..475..295L}. The project
provided feedback to the IVOA on the SAMP protocol, allowing for
inclusion of a full SED into a single file extension, to TOPCAT
\citep{2005ASPC..347...29T,2011ascl.soft01010T} for better support for SED plots, and inspired work toward a Virtual
Observatory Data Model Language (VODML) by lead Iris developer O.\ Laurino. 


\subsection{Scalable Cross-Comparison Service}
\begin{figure}[tb]
\centering
\includegraphics[width=0.9\columnwidth]{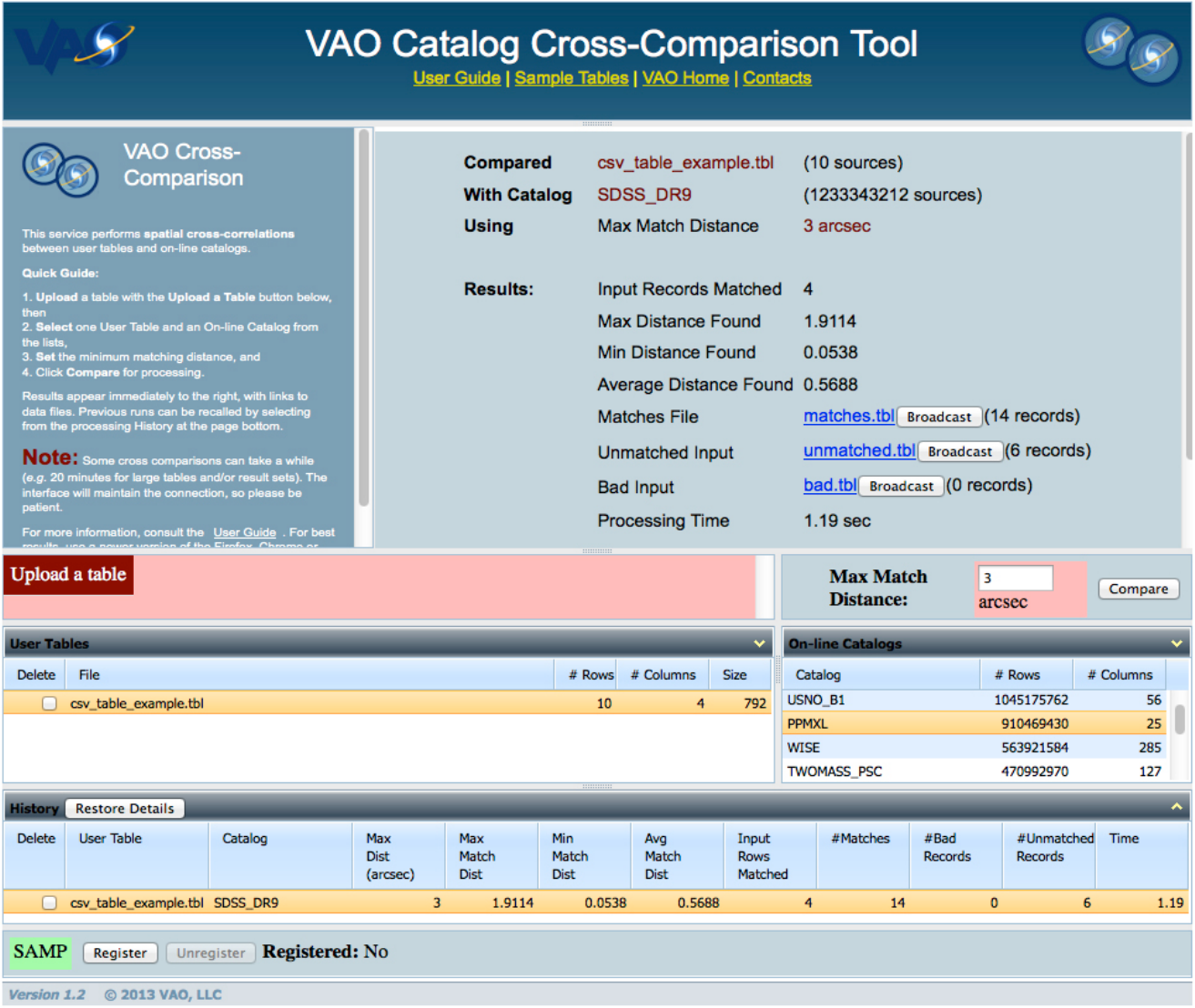}
\caption{The Scalable Cross Comparison Service.  Shown are results for a user-uploaded table cross-matched with the Sloan Digital Sky Survey Data Release~9 catalog.}
\label{fig:SCCS}
\end{figure}

The Scalable Cross Comparison (SCC) Service performs fast positional cross-matches between an input table of up to 1 million sources and common astronomical source catalogs for a user-specified match radius. The service returns a list of cross-identifications to the user. The output is a composite table consisting of records from the first table, joined to all the matching records in the second table, and the angular distance and position angles of the matches (Figure~\ref{fig:SCCS}).

The first release of the Scalable Cross Comparison Service was in~2012
January and was supported with three upgrades over the next
1.5~years. The indexing schemes that support large catalog cross-matching were
provided by the Infrared Processing and Analysis Center (IPAC) and
later adapted to the Wide-field Infrared Survey Explorer (WISE) and \textit{Spitzer} projects.

\subsection{Time Series Search Tool}\label{sec:time}
The Time Series Search Tool finds and retrieves time series data from three major archives and analyzes them with the NASA Exoplanet Archive periodogram application. The application was a prototype developed to demonstrate that the IVOA standards of the time-series protocol and data model met the needs of such a tool. The development of the Time Series tool ended after the first VAO re-plan.

\subsection{Lessons Learned}

{The VAO science applications group was distributed across
multiple institutions, and \cite{2012SPIE.8449E..0IE} described the
management strategy this group.  A key element of developing
successful applications amongst a distributed group is managing
unknowns.  As the VAO science application lead might be unaware of the entire set of tasks assigned to an individual outside of the VAO efforts, coordinating task assignments and making organizational material and schedules easily available was important.}

{The VAO implemented a relatively lightweight process, tracked in
a Wiki-based environment, in order to focus the distributed team on
the requirements, design, and implementation of the applications.  In
addition to the developers themselves, a science stakeholder was
assigned to each application and was key to bringing the view of the
user to the development process. The stakeholder provided
requirements, developed science use cases, handled technical
questions, advised on development priorities, and performed unit
tests.  The use cases drove development and provided an opportunity to
assess priorities and make course corrections.  Internal product
deliveries provided a test and assessment loop, and incremental
releases (rather than one big software release) ensured that
development was progressing as expected.  A team lead managed
priorities, schedule, and communication within the group.}

{Frequent communication was essential to ensuring that issues
were resolved quickly and the team was working toward a common vision.
The distance gap of distributed teams needs to be managed diligently.
The VAO Wiki provided easy-to-access project information so that a
team member could resume work quickly if he or she were sidetracked
due to external project responsibilities.  This process enabled the
group of developers working on a project, at a distributed set of
institutions and working on a part-time basis, to perform their tasks
and collaborate efficiently \citep{2012SPIE.8449E..0IE}.}

\subsection{Community Developments}\label{sec:community}

{%
During the course of the \hbox{VAO}, there were diverse, community-led
efforts to develop VO software.  (In some cases, these efforts started
during the NVO era, but continued into the VAO project.)  Often these
involved VAO personnel, either in the role of ``consultants'' or who
were engaged through their work on other projects.}

{%
Examples of such community-led software efforts include 
VOEvent, a protocol for notifications or ``alerts'' from and between
observatories \citep{2006AN....327..775W};
Montage, a user-controlled tool for generating science-quality image
mosaics \citep{2003ASPC..295..343B}; and 
seleste,\protect\footnote{
\texttt{http://cda.cfa.harvard.edu/seleste/}}
a tool designed to provide uniform access to distributed VO databases.
}

\section{Standards and Infrastructure}\label{sec:SandI}

The core of the VAO program was the development of software to support the IVOA standards for discovery and access to distributed data.  Key components of the VAO infrastructure include the resource registry (the collection of metadata describing on-line data collections and services), 
the data access layer protocols (images, spectra, tables, databases) and their validation tools, a distributed authentication service (``single sign-on''), and applications programming interfaces either built-in to existing software packages or available stand-alone that allow researchers to develop their own VO-enabled scripts.  Much of the VAO infrastructure is now incorporated into the data services of major data centers using VAO-provided software libraries.

\subsection{The VAO Infrastructure in Context}\label{sec:voarch}

\begin{figure}[tb]
\centering
\resizebox{\hsize}{!}{\includegraphics{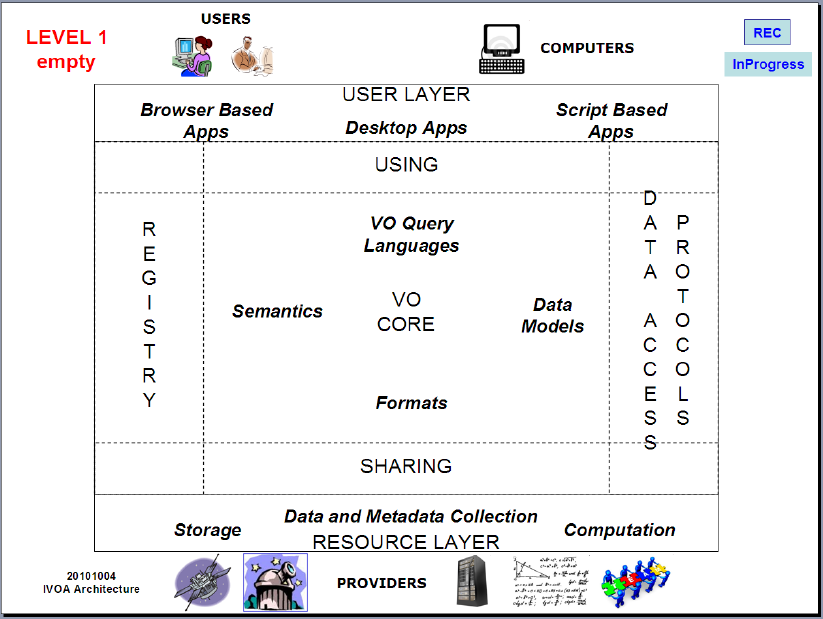}} 
\caption{
{Virtual Observatory architecture \citep{IVOA_Arch}}.
Users appear at the top of
the figure and data providers and computational resources are at the
bottom, connected by the VO bridge.  The VO bridge itself
comprises the registry of data providers and data services, the data access protocols for discovering and retrieving data, and the core infrastructure of query languages, data models, data formats, and semantic definitions.}
\label{fig:VOArchitecture}
\end{figure}

Figure~\ref{fig:VOArchitecture} shows the {VO architecture \citep{IVOA_Arch}}.
In this diagram, the VO infrastructure serves as a bridge between data providers and users, and that bridge is supported by standards. On the provider side, data is connected into the infrastructure through standard services that present that data in terms of standard data models. On the other side, users are connected to the infrastructure via generic tools that understand the VO standards. Tools are no longer tied to a single archive, but rather can talk to any and all archives that speak the common VO language.

\begin{figure}[tb]
\centering
\resizebox{\hsize}{!}{\includegraphics{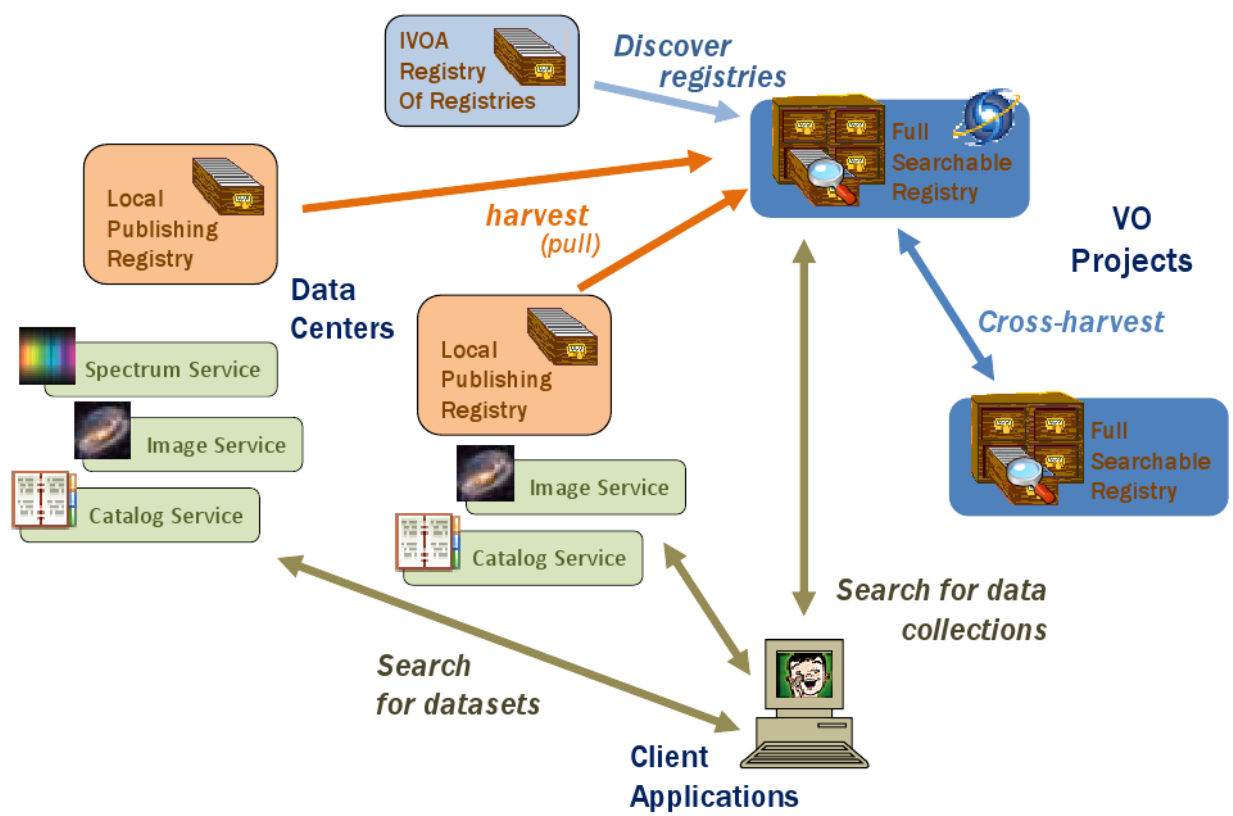}} 
\caption{Data discovery in the Virtual Observatory.}
\label{fig:Registries}
\end{figure}

Providing the ability to discover and access data of interest is a significant
motivation for the structure of the VO architecture.
Figure~\ref{fig:Registries} illustrates the discovery framework. 
Registries represent the first step for data discovery in this
framework.  A registry is a database containing descriptions of data collections and services available in the VO
\citep{2014A&C.....7..101D}.  Conceptually a VO registry is similar to
a ``name server'' for domain name service (DNS) on the Internet
\citep{RFC1034}.

There is no single master or central registry; however, there are
registries called full searchable registries that aim to have
descriptions of all the data collections, archives, and service
providers known to the VO from around the world. This type of registry
can populate itself through a process known as harvesting; it starts
by contacting a special ``boot-strapping'' registry (run by the VAO)
called the Registry of Registries that will return to it all of the
other known registries in the VO ecosystem. In order for a new
registry to enter the \hbox{VO}, it must be registered with the
Registry of Registries.

Most of the registries within the VO 
are publishing registries. A registry of this type is
typically run by a data center that uses it to advertise the data
collections and services that it offers to the \hbox{VO}. The full
searchable registry contacts each of the publishing registries
and pulls descriptions of all the data collections and services
provided by the data center. At this point the full searchable
registry is populated with descriptions of all of the resources
known to the \hbox{VO}. Periodically it will re-query the other
registries to obtain any new resources or other changes since the last harvest.

With an up-to-date full searchable registry available to it, a client
application (e.g., \S\ref{sec:SciApps}) can discover any data known to the VO. It starts by asking the registry for a list of collections and services from each of the data centers that might have data relevant to the user's science question. Most of the services will be standard data access services for finding and downloading images, spectra, or catalog information from a particular archive or collection. The application can then send a query to all of the matching services to get back lists of available data sets. By browsing the returned metadata for these data sets, the user can choose which data sets to download. 

\subsection{VAO and the IVOA}

Much of the work the VAO conducted in advancing standards was through
engagement with the \hbox{IVOA}. The role of the IVOA is two-fold:  first, to coordinate the efforts of all of the VO projects around the world, and second, to serve as a standards body for establishing VO interoperability.

From the IVOA's beginnings, the NVO and VAO were leaders in shaping
the VO's global architecture and the standards that enable it,
reflecting the significant data holdings of U.{}S.\ institutions.
NVO/VAO staff members served as chairs or vice-chairs of key IVOA
working groups (\ref{app:ivoa_lead}). The impact of this leadership is
also seen in the standard documents; most of the IVOA recommendations
across all of the areas of the VO have featured NVO/VAO team members
either as first authors, secondary lead authors, editors, or major
contributors (\ref{app:standards}).

The VAO produced many of the key reference
implementations---software that demonstrates a standard in action and
proves its viability.
{During the NVO era, there was a vigorous international debate
regarding the character of the VO Registry and whether it should be
relatively ``coarse-grained'' or ``fine-grained,'' in terms of the
amount of detail stored in the VO Registry.  (See below.)}
The NVO created the first implementations of
registries with several different architectures.  The VAO was
instrumental in demonstrating data access services through software
packages like DALServer (\S\ref{sec:dalserver}) and TAPServer
(\S\ref{sec:tapserver}).

The NVO/VAO led the IVOA in the development of service
validators. A validator is an application that checks whether another
service is compliant with VO standards.  A validator performs this
check by sending a series of queries to a VO service and
examining the response to assess whether it follows all of the
rules and recommendations described in the standard. The NVO developed
the first validators in the IVOA to assist data providers, allowing
them to check their data access services and fix any problems before
publishing them to the \hbox{VO}. These NVO validators quickly became
critical pieces of VO infrastructure and were continued by the
\hbox{VAO} (\S\ref{sec:valid}); and other projects joined in to contribute validators for other service standards.

In other areas, though, the VAO benefited from international
developments.  Not only did the VAO benefit from technical comments,
there were multiple occasions in which the VAO could produce a library
or tool, thereby ultimately more rapidly because some of the initial
development had been done by international partners (e.g., the
development of the single sign-on capability, initially developed by
Astrogrid).

\subsection{The Registry}\label{sec:registry}

As described above, a VO registry is a database containing descriptions of data collections, archives, services, and other resources, and it represents the first step in data discovery.  The NVO/VAO established itself as an early leader in the area of registries. In addition to creating some of the first registries, the VAO operated the Registry of Registries (RofR) on behalf of the \hbox{IVOA}. The RofR allows searchable registries to bootstrap their collection of resource descriptions. 

{%
The NVO and its IVOA partners developed several different types of registries with several different implementations.  There was considerable debate over the registry design, and whether it should be ``fine-grained'' or ``coarse-grained''.  A fine-grained registry contains detailed metadata about the datasets available at a VO resource (for example, it might contain the right ascension and declination of all observed positions in an archive).  A coarse-grained registry would only contain information about the general sky coverage of an archive.  The advantage of a fine-grained registry is that one need not query distributed resources explicitly to determine if they have data of interest, whereas with a coarse-grain registry data discovery is a two-step process.  The problem with a fine-grained registry, however, is that many data collections are dynamic, so that any metadata cache has to be updated continuously.  Also, the structure of a fine-grained registry will necessarily be much more complicated, and harvesting of metadata between fine-grained registries could easily become inefficient.  Despite the efficiencies for search and discovery offered by fine-grained registries, the VO currently operates with coarse-grained registries.  The VAO consolidated support around the coarse-grained, full searchable registry service at the Space Telescope Science Institute.  There are ongoing efforts to build a fine-grained registry for mostly static data collections.}

\subsubsection{VAO Directory Service}

\begin{figure}[tb]
\centering
\resizebox{\hsize}{!}{\includegraphics{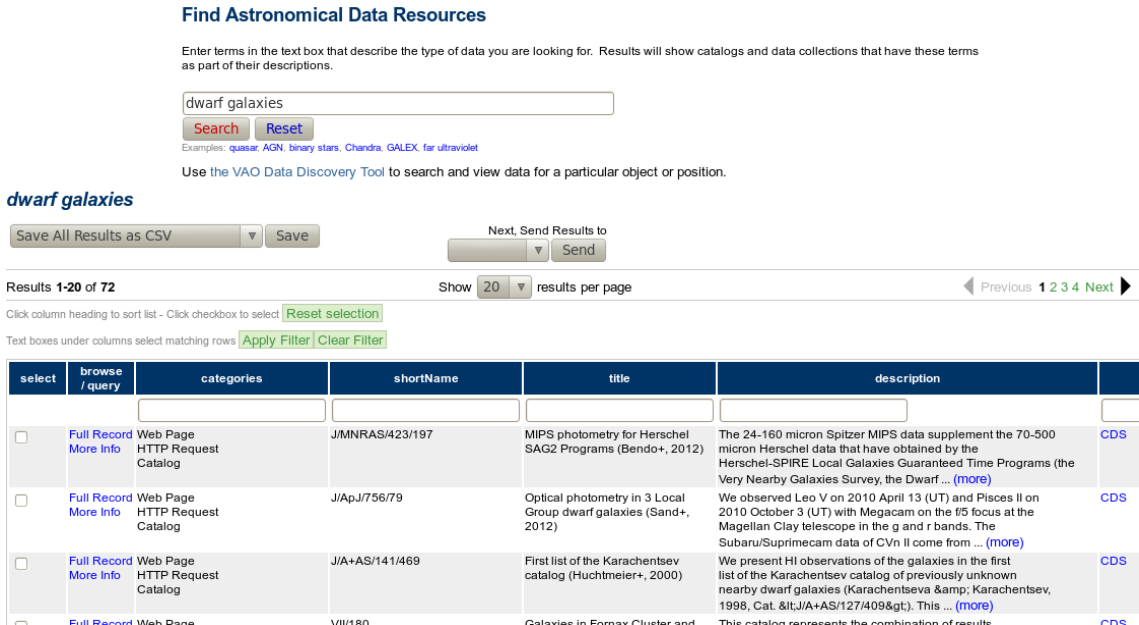}} 
\caption{Results from a search query submitted through the VAO Directory Service.
The directory service results allow one to browse the descriptions, filter results, download matching descriptions in VOTable format, and when the resource is a standard data access service, even send it a sky position-based query.}
\label{fig:DirectoryService}
\end{figure}

As part of the VAO's production registry, a Web-browser-based front
end called the Directory Service\footnote{
\texttt{http://vao.stsci.edu/directory}}
was provided.  This tool is particularly useful for discovering collections and services related to a topic. By entering keywords into the search input box, the tool will return a list of resources whose description contains those keywords (Figure~\ref{fig:DirectoryService}).

\subsubsection{Registry Upgrades}

Over the last two years of the VAO project, an updating program was
conducted to overhaul the underlying registry database and update it
to support the latest IVOA registry metadata standards. This overhaul
was also necessary to support a new standard for searching
registries. This new standard leverages an existing IVOA standard for
querying complex databases called the Table Access Protocol
(TAP), for which client software already exists. (The TAP standard did
not exist when the first registry search interfaces were
standardized.)  Completing this upgrade was
critical to maintaining the registry in the eventual post-VAO era.

A final effort conducted in the VAO project was to complete registry
curation activities aimed at improving the descriptive content of the
registry. In particular, a specific approach was implemented to
registering resources intended to make registry searches more
effective and their results less confusing. This approach has recently
been accepted as a best practice by the IVOA Registry Working
Group. The VAO curation work started with an inventory of existing
resources by publisher, followed by developing a set of
recommendations for improving the resource descriptions that brings
them into line with the best practice.  The new registry resource
publishing tool (described below) will be instrumental in communicating these recommendations to the publisher.

\subsubsection{Publishing Registries and the Resource Publishing Tool}

A publishing registry is the vehicle for making a resource available
to the \hbox{VO}. In particular, it can create new descriptions of
resources and share them with the rest of the VO through the
harvesting process. A data center, which may curate a number of data
collections and offer a variety of services to access them, may
operate their own publishing registry. Because such a registry does
not need to serve end users directly, operating one is much simpler
than running a searchable registry. During the NVO project, the
VORegistry-in-a-Box product was developed that provides a simple but
compliant publishing registry implementation through which a data
center can maintain its own resource descriptions in-house. This
product is still in production use within the VO (including by the
Registry of Registries), and the VAO continued its support.

A searchable registry can also support the publishing function, which the VAO Registry at STScI does. In particular, it
maintains resources descriptions on behalf of data providers who only
have a few resources to share, relieving them from having to run their
own publishing registry. In order to enable this feature, the VAO
created the Resource Publishing Tool, a browser-based application that
allows a data provider to create and share resource descriptions
through the VAO Registry. It features a guided interface that steps a
data provider through the process of describing a resource, prompting
for metadata along the way.  The tool also can check for the validity of values as they are entered, alerting the user of any problems. Draft descriptions can be saved for updating and publishing later, and already-published resource entries can be updated with this tool. Various techniques are used to minimize the amount of typing required to create a useful resource description. While the VAO Registry will share records created through this tool, the descriptions are considered ``owned'' by the user. Thus, to control access, the publishing tool uses the VAO Single Sign-On Services (described below).

\subsection{Data Access}

Standard services that allow users to find and access to data from an
archive are part of the VO architecture known as the data access layer
{(\hbox{DAL}; Figure~\ref{fig:VOArchitecture})}.
In the VO architecture, there is a standard service for each
type of dataset; e.g., the Simple Image Access protocol \citep[\hbox{SIAP}, ][]{IVOA_SIAP} enables
discovery and downloading of images from an archive, and the Simple
Spectral Access \citep[\hbox{SSAP}, ][]{IVOA_SSAP} protocol enables
access to spectra.
{In this section, we describe the four different ``toolkits'' or
new protocols that the VOA developed for improved data access within the \hbox{VO}.}

\subsubsection{DALServer}\label{sec:dalserver}
 
In order to help data providers share their data collections through
standard VO services, the VAO created the DALServer Toolkit, a
Java-based software package.  When first developed as part of
the NVO project, it served as a platform for developing reference
implementations of standard VO services (like SIA and SSA) that
demonstrated features of the standards.
{%
At about the same time, both Astrogrid and ESO were developing data
access toolkits, and some of this development fed into the VAO concept.}

During the VAO's final year, specific efforts were made to enhance the
toolkit for use directly by data providers; this effort was considered
``productization,'' as it focused on making the toolkit easier to use.
The focus was on a simple class of use cases in which a small data
provider had a simple catalog or a simple collection of images or
spectra that they wished to share. By just editing configuration files
and running a few scripts, the provider could deploy fully compliant
VO services with no programming required. For more complicated
situations, such as for a data center that might already operate
custom data access services through their own data management system,
they could use the underlying DALServer Library application
programming interface (API) to adapt the VO services to their local
infrastructure.

The first production release of the DALServer provided support for the
four ``simple'' standards for data access recommended by the
\hbox{IVOA}:  namely, Simple Cone Search \citep[\hbox{SCS}, for simple
position-based querying of object and observation catalogs, ][]{IVOA_SCS}, Simple
Image Access Protocol (\hbox{SIAP}, for finding images), Simple Spectral Access
Protocol (\hbox{SSAP}, for finding spectra), and Simple Line Access
Protocol \citep[\hbox{SLAP}, for finding rest frequencies for spectral line emissions, ][]{IVOA_SLAP}.  Toward the
end of the VAO project, DALServer was extended to operate on
multidimensional data sets (\S\ref{sec:multid}).

\subsubsection{TAPServer}\label{sec:tapserver}

The Table Access Protocol (TAP) is an IVOA standard for querying
complex catalogs that may be made up of several tables (e.g., the
2MASS catalog). When a TAP service is connected to a catalog, users
can create complex, SQL-like queries that can join metadata from
several tables. Such queries are critical for mining very large
catalogs. Not surprisingly given its power and flexibility, a TAP
service is one of the more complex IVOA standards to implement. To
make deploying a TAP service easier, the VAO created the TAPServer toolkit.

Like DALServer, TAPServer is configuration file driven. That is, with no
programming required one can wrap the toolkit around a collection of
tables in a database and deploy it as a service accessible to the
\hbox{VO}.  Because of the VAO close-out schedule, only a limited
amount of development could be completed, and there was no effort
toward the ``productization'' of TAPServer. However, the code is
included in the VAO Repository and available for community use. Some
post-VAO targeted deployments are planned. For example, it will be
deployed at the National Center for Supercomputing Applications (NCSA)
to expose the Dark Energy Survey Source Catalog.  In turn, DES
scientists will be able to analyze the catalog using the 
seleste\footnote{\texttt{http://cda.cfa.harvard.edu/seleste/}}
TAP client, a tool that allows users to form complex queries with little or no knowledge of \hbox{SQL}.

\subsubsection{Service Validators}\label{sec:valid}

During the VAO project, service validators originally developed during the NVO project were continued and expanded.  These validators have a web browser interface that allows a data center to enter a service access URL and test the serviceÕs compliance with the appropriate standards; the result is a listing of errors, warnings, and recommendations for improving the service. These validators share a common Java-based toolkit platform called DALValidate. They also support a programmatic interface that allowed VAO Operations to automatically test VO services. (The VAO Operations team also engages other validators developed outside of the \hbox{VAO}.)  Supported validators include those for Simple Cone Search, Simple Image Access, Publishing Registries, and VOResource records.  The DALValidate software is available through the VAO Repository.

\subsubsection{Image Cube Access}\label{sec:multid}

An emerging suite of telescopes is or soon will be generating
multidimensional data (often termed ``image cubes'').  The most
general data set, produced by an instrument measuring photons, would
be $[I(\alpha, \delta, \nu, t), Q(\alpha, \delta, \nu, t),
U(\alpha, \delta, \nu, t), V(\alpha, \delta, \nu, t)]$, where we have described
the polarization properties by the Stokes parameters $(I, Q, U, V)$
and each polarization can be a function of position on the sky
$(\alpha, \delta)$, frequency~$\nu$ (or equivalently wavelength~$\lambda$ or
energy~$E$), and time~$t$.  Radio interferometers have naturally produced
such multidimensional data sets for some time, and the commissioning
of the Jansky Very Large Array (JVLA) and the Atacama Large
Millimeter/submillimeter Array (ALMA) is making such data sets much
more common.  X-ray telescopes have, for some time, been generating
data that can be considered to be extremely sparse image cubes.  The introduction
of integral field units (IFUs) both for ground-based telescopes and
eventually for the \textit{James Webb Space Telescope} is making these
data more common at visible and infrared wavelengths.

Providing discovery of and access to multidimensional data was taken
up as a key project upon endorsement of the VAO Board.  The VAO
also helped stimulate interest in multidimensional data within the
\hbox{IVOA}.  In the \hbox{IVOA}, it was recognized that although SIAP could support image cubes in a
limited way, it lacked some of the metadata support and data access
mechanisms needed to support the cubes being produced or soon to be
produced.  From the VAO perspective, not only would discovery and
access to multidimensional data advance a new capability in the
\hbox{VO}, it might also to engage the radio astronomy community more in VO activities.

The VAO produced an early prototype service that demonstrated a number
of the key capabilities needed in a new standard for image cube
discovery and access. This demonstration was instrumental for mapping
out the strategy for an SIAP Version~2 \citep{IVOA_SIAPV2}.  In particular, the
necessary standardization was broken down into three independent
components:  (1)~the Image Data Model defines the semantic labels used
to describe image cubes; (2)~these labels are used by the SIAPV2
standard to annotate image search results; and (3)~the Access Data
standard defines how one can request cutouts or other transformations of image cubes.

While active in the development of the standards within the
\hbox{IVOA}, the VAO continued prototyping access to image cube
data. To ensure that the standards served the needs of real providers
of image cubes, we established a collaboration with the National Radio
Astronomy Observatory (NRAO). Our joint goals were first to create a
real functional image cube access service based on the emerging SIAV2
draft serving real data from NRAO instruments, and second, to provide
a useful architectural design along with software to support active
archive operations. In this collaboration, NRAO provided the VAO
project with requirements and use cases. NRAO desired an image service
that could simultaneously provide data to both internal and external
clients. One key client is the Common Astronomy Software Applications
\citep[\hbox{CASA}, ][]{2008ASPC..394..623J,2011ascl.soft07013I} Viewer that needs to request small
visualizable parts of a larger cube. In return, the VAO project
provided NRAO with general purpose software to both deliver data over
the network to clients. The DALServer product (\S\ref{sec:dalserver}) was extended to provide server-side support for SIAV2, and VOClient (described below) was extended to support the client.  In the spring of~2014, NRAO, using VAO-provided software, successfully demonstrated a service that provides access to image cube data, including image cutouts. This service allows their archive and CASA Viewer developers to test against a functional service.

\subsection{Data Sharing}

{
A common interest among astronomers is making their data available to
their colleagues.  Data sharing can be essential part of a project in
which team members are in different institutions, or it can be for
legacy reasons to enable data re-purposing (using the data for studies
not originally envisioned when the data were acquired) or for ensuring
replication.  While there are institutional data centers, both in the
U.{}S.\ and internationally, there are also so-called ``long-tail''
data, the many small collections of data products that are typically
associated with published papers. Such data products tend to be highly
processed by individual astronomers and are not typically available
from traditional observatory or project archives.
}

The second key project conducted during the close-out plan was more
exploratory in its full scope (though it supported an important
end-user application). 
The VAO sought to understand how these products could be published to
the VO in a low-effort way; in order to enable such access, the focus
was on integrating data sharing and publishing into the overall scholarly publishing process
which starts even before the first draft of a paper.  Two products
were developed.

\subsubsection{SciDrive}

\begin{figure}[tb]
\centering
\resizebox{\hsize}{!}{\includegraphics{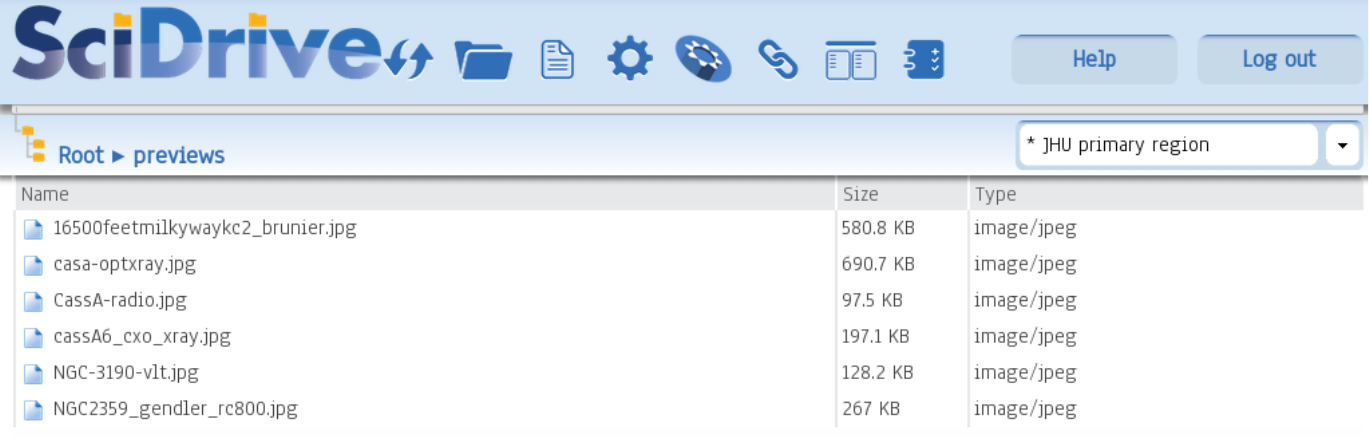}} 
\caption{The SciDrive file browser, viewed via a web browser.}
\label{fig:SciDrive}
\end{figure}

SciDrive is a Dropbox-like cloud storage application intended for use in
scientific research \citep{2014ASPC..485..465M}. It {was inspired by the SDSS MyDB
\citep{2004ASPC..314..372O} and AstroGrid MySpace
\citep{2004ASPC..314..330D} developments}, and it is based primarily
on the OpenStack\footnote{ 
\texttt{http://www.openstack.org/}}
software (in particular the OpenStack Swift component for object
storage).  It can be accessed from a web browser in which the user is presented with a view of a personal hierarchical directory space where one may save files by dragging-and-dropping file icons into the web page interface (Figure~\ref{fig:SciDrive}). Also available is a desktop client that can (like Dropbox) monitor a local directory and automatically upload files that are moved into it. As many researchers already do with Dropbox, SciDrive can be a simple platform for sharing data within a research group; it provides a secure means to share read-write access to a collection within a restricted group or to send one-off permissions (read or read/write) to individuals. One difference from commercial storage providers is SciDrive's ability to scale to larger collections than with the typical free versions of storage.

SciDrive supports the VOSpace~2.0 interface, the IVOA
standard for managing third party data transfers \citep{IVOA_VOSpace}. This capability allows a
user to seamlessly move files between different SciDrive instances (or
other VO-compatible storage systems) located around the network.  This
feature is important for new VO capabilities in which web-based tools allow users to save application outputs to their personal space in the cloud. These outputs could be reloaded later into the tool for further analysis (e.g., as a ``favorite'' starting point) or loaded by other tools for synthesis with other data and analysis. A current example of this is the use of SciDrive with the Sloan Digital Sky Survey (SDSS) CasJobs: a SciDrive user can configure a directory to automatically detect uploaded table files and load them into the SkyServer database so that it can be correlated with the SDSS catalog.

The CasJobs connection highlights another unique feature of SciDrive:
it supports plugins that enable special handling of certain types of
data.  This application is therefore a possible platform for
publishing data by individual scientists and research groups. There has
been experimentation with plugins that automatically extract the metadata
from files that is needed to expose data to the
\hbox{VO}. With such a feature, a research group could use SciDrive to organize a collection of data for publication. When the collection is ready for release a simple
press of a button would expose the data publicly; the metadata would be automatically loaded into a database and the collection would be made available through standard IVOA services (e.g., using DALServer).

Completing this vision to a working implementation was beyond the
scope of the VAO Project; nevertheless, operations and development at
Johns Hopkins University (JHU) of the SciDrive platform continues (with
NSF support from the Data Intensive Building Blocks program). Furthermore, VAO partners JHU and the National Center
for Supercomputing Applications (NCSA) are collaborating on the emerging, community-driven initiative called the National Data Services (NDS) Consortium, which aims to address data publishing across all research fields. As the publishing scenario described above is much like one being discussed in the NDS community, we expect the development of SciDrive as a publishing platform to continue beyond the VAO project.

\subsubsection{Single Sign-On Services}\label{sec:sso}

In order to restrict access to the user's personal space, SciDrive
uses the VAO Login Services for authentication \citep{2012ASPC..461..423P}.  These services were created so that VO users
could have a single login to connect any VO-compatible service or
portal even when they are managed by different organizations. More
than the simple convenience of a single login, a federated login
system allows a user to access their proprietary data from one data
center using analysis tools from another data center. A participating
organization can choose to support VAO logins either as its primary
identity or as an augmentation of its local authentication system.

{Inspired by initial developments by Astrogrid for the single
sign-on capability in an astronomical context,} the VAO federated login
is built on the OpenID standard\footnote{
\texttt{http://openid.net/}}
that is in broad use across the Internet. Associated with it are all
the usual services that help users manage a login: the ability to
reset forgotten passwords, edit the user profile, etc.  The VAO Login service also leverages an OpenID feature for sharing user information with a portal in a privacy-conscious way; this can make registering users with a portal faster and simpler. One less common feature that is important for VO applications is the ability for transparently delivering X.509 certificates to the portal. This allows a portal to access private data at another site on the user's behalf. While the service requires the user's permission to do this, it is worth noting that the user never handles the certificates directly.

The development of the VAO Login service resulted in two release software products. First, VAOSSO provides the user identity server that powers the VAO services. This software can be configured either to run as a mirror of the VAO service (for high availability) or as a completely independent service.  Second, VAOLogin is a toolkit that helps portal developers add support for VAO Logins.

Current applications using the VAO Login Services include SciDrive,
the VAO Registry's Resource Publishing Tool, and the VAO Notification
Service. The National Optical Astronomy Observatory (NOAO) Data
Archive, which currently supports the predecessor NVO Login Service, is migrating to use of the VAO Login Services to augment their own local authentication system.

\subsection{Virtual Astronomy on the Desktop}

A key initiative of the VAO Standards and Infrastructure program was
to make VO capabilities more available from a user's local
machine. Not only was the goal to make VO capabilities integrated into
both new and existing desktop applications, the VAO Project sought to
deliver that power directly to scientists through custom scripts that
they can create to conduct their research.

Because of its growing popularity as a scripting language for
scientific research, Python\footnote{
\texttt{https://www.python.org/}}
was a major focus of our scripting support, {following upon the
example set by AstroGrid's python package}.
Further, we enabled all VO-enhanced applications and scripts running
on the desktop to work together using the Simple Application Messaging
Protocol \citep[\hbox{SAMP}, ][]{IVOA_SAMP}, the IVOA standard that
allows desktop and Web applications to exchange data.

\subsubsection{VO-Enhanced Image Reduction and Analysis Facility (IRAF)}
The first VAO product supporting VO on the desktop was a VO-enhanced
version of IRAF \citep{1986SPIE..627..733T,1993ASPC...52..173T,1999ascl.soft11002N}
developed by M.~Fitzpatrick (NOAO). This
included some general IRAF infrastructure enhancements including the
ability to load data from arbitrary URLs as well as support for
loading data in VOTable format. With these two capabilities, a suite
of tasks was added to take advantage of VO services; these included
an object name resolver, the ability to search the registry to find
archives and services, the ability to search individual archives or
catalogs, and the ability to download discovered data products. SAMP
support was also added so that IRAF could send data to other non-IRAF
tools running on the desktop; for example, images could be sent to
Aladin \citep{2000A&AS..143...33B} and catalogs to TOPCAT for visualization.

\subsubsection{VOClient}
This downloadable product provides direct access to VO services outside of a Web browser. The first VOClient release featured a suite of command-line tools that enables interactive use from UNIX/Linux shell; they can also be used to create customized shell scripts. The capabilities provided by these tools include discovering archives and catalogs via the VAO registry, searching individual archives for images and spectra, downloading discovered data across multiple archives, searching catalogs by position, resolving object names to sky positions, and sending data to other desktop tools (via SAMP).

The second release of the VOClient package focused more on the
underlying set of core C libraries. These libraries can be used
directly to add VO capabilities to C and C++ applications (as was done
for the NRAO CASA Viewer). These libraries are intended to be the
basis for bindings to other languages, such as Python and
Perl.\footnote{
\texttt{https://www.perl.org/}}
The Python bindings in particular were a focus of the second release (which featured a common API with \hbox{PyVO}, described below). Finally, the second release featured a task framework that enables easy integration of legacy software, making it callable from Python.

\subsubsection{PyVO}
\begin{figure}[tb]
\centering
\resizebox{\hsize}{!}{\includegraphics{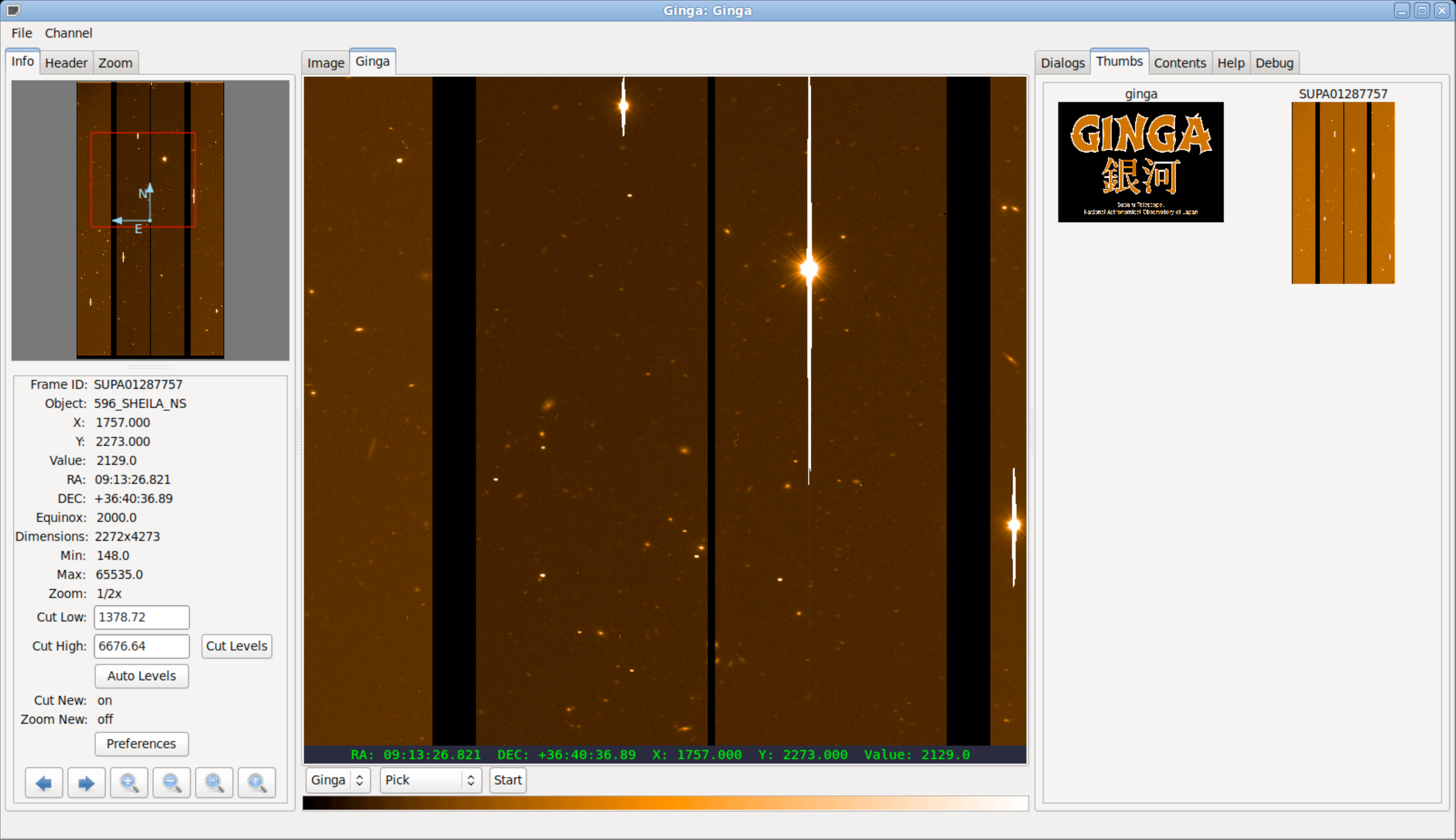}}
\caption{Ginga Image Browser from the Subaru Telescope showing a VO plugin powered by \hbox{PyVO}.
The rightmost panel represents a plugin that allows users to download images and catalogs from the VO for display and overlay in the viewer.}
\label{fig:Ginga}
\end{figure}

This downloadable product represented a parallel effort to support
Python with a slightly different focus. Through our community
engagement, we found that many Python users prefer to use a
pure Python implementation of a VO library, which PyVO provides,
{as opposed to a mixture of Python and Unix system commands.}
As for VOClient the audience is two-fold, the first being developers
who want to integrate VO capabilities into their own Python
applications. As an example, Figure~\ref{fig:Ginga} shows the Ginga
image browser, developed for the Subaru Telescope, to preview
observatory images \citep{2013ASPC..475..319J}.  Downloading of images
and catalogs was an additional functionality added to the Ginga image
browser using the PyVO python module.

PyVO was also aimed at the growing community of research astronomers
using Python to create custom scripts to carry out their research and
analysis. In fact, PyVO is built on top of the widely used Astropy
package \citep{2013A&A...558A..33A}, an integrated set of astronomically-oriented modules. This allows users to discover and download data and process and analyze it with the robust capabilities of Astropy. This combination is an important key to doing VO science at a large scale, as it becomes very easy to apply common processing to a vast array of data either from a single survey or from distributed collection. It also becomes possible to continuously monitor the evolving holdings of an archive or the VO in general as new data sets are added.

The first evaluation version of PyVO was released 2013. As this
release date was close to the end of the VAO Project, we wanted to
ensure further use and development of PyVO beyond the Project's end.
Accordingly, we explicitly employed a strategy to build a community
around the PyVO package. First, GitHub\footnote{%
\texttt{https://github.com/}}
was used to provide a web-based
code repository for future community contributions. This approach has enabled important contributions from users outside of the VAO Project;  as of this writing, there are 22 issue submissions from seven external users and seven code submissions from four external users. The other part of the strategy was to establish a strong tie to the Astropy community, which is quite large and active. (In fact, this tie is responsible for much of the external participation via GitHub.)  To this end, we applied for and were given status as an Astropy ``affiliate package.''  This connection also allows PyVO to become a proving ground for migrating addition VO capabilities into Astropy.

\section{Operations}\label{sec:ops}

The VAO operations effort addressed two primary goals. The first was
to enable science use of the \hbox{VO}, in the sense of being an ``operational observatory,'' with a focus on
the VAO-developed interfaces but not exclusively. Tools must work, should work consistently, and when problems arise they must be swiftly resolved. The second goal was to enable the services needed internally for the activities of the VAO itself. VAO personnel needed reliable access to the tools needed for software design and access, user support, testing, configuration management, bug tracking, and so forth.

The VAO provided a number of science services and tools directly to
the scientific community (\S\S\ref{sec:SciApps} and~\ref{sec:SandI}):
its home web site, a data portal and cross-corrlation tool, the Iris
SED tool, downloadable VO libraries for use by clients and servers,
and cloud storage and secure access protocols. Internal services
included the VAO infrastructure: the JIRA ticket system, a Jenkins
testing service, SVN code repository, a YouTube channel,
a blog, and mailing lists.  The VAO also supported the IVOA Web site and document repository; these were transferred to international partners in Italy and India.  The VAO software repository\footnote{
\texttt{https://sites.google.com/site/usvirtualobservatory/}}
was established to ensure that VAO-developed resources are available indefinitely.

VAO services are supported by member institutions of the VAO with
significant resources hosted at each of our sites: the Smithsonian
Astrophysical Observatory, \hbox{JHU}, \hbox{MAST}, \hbox{HEASARC}, \hbox{NRAO}, \hbox{NOAO}, Caltech, and IPAC (IRSA and NED). Most recently the software repository has used free Google cloud-based services. Elements are distributed across the country and the Internet.

Supporting such a distributed system posed (and will pose) special
operational concerns.  Especially for its science users, the VAO
worked to ensure that elements were seen as a coherent whole: science tools need to be available at a common location, forms should have consistent look-and-feel, and everything should be clearly visible through a consistent web presence even when the web sites are on various servers.

All elements were continuously monitored and a responsible party
identified for each so that issues could be rapidly and decisively
addressed. The operations staff met frequently (in weekly
telecons) and operational issues were rapidly escalated using an
internal issue tracking software to whatever level was needed to
ensure that they received the needed visibility.

\subsection{Service Monitoring}

All VAO services were monitored hourly and a database of all tests was
continuously updated. Each service was tested to ensure not only that
the service was operational, but also that it responded sensibly to
some simple request. When services failed a test, they were retested
15~minutes later. If the second test also failed, a message was automatically sent to the responsible parties and to the VAO operations monitor.

A web site was available giving the current status of all operational
services, and the VAO home site reflected the operations status of VAO
science services so that users were immediately informed if there was an issue. Statistics were collected in and reported in biweekly periods.

\begin{figure}[tb]
\centering
\resizebox{\hsize}{!}{\includegraphics{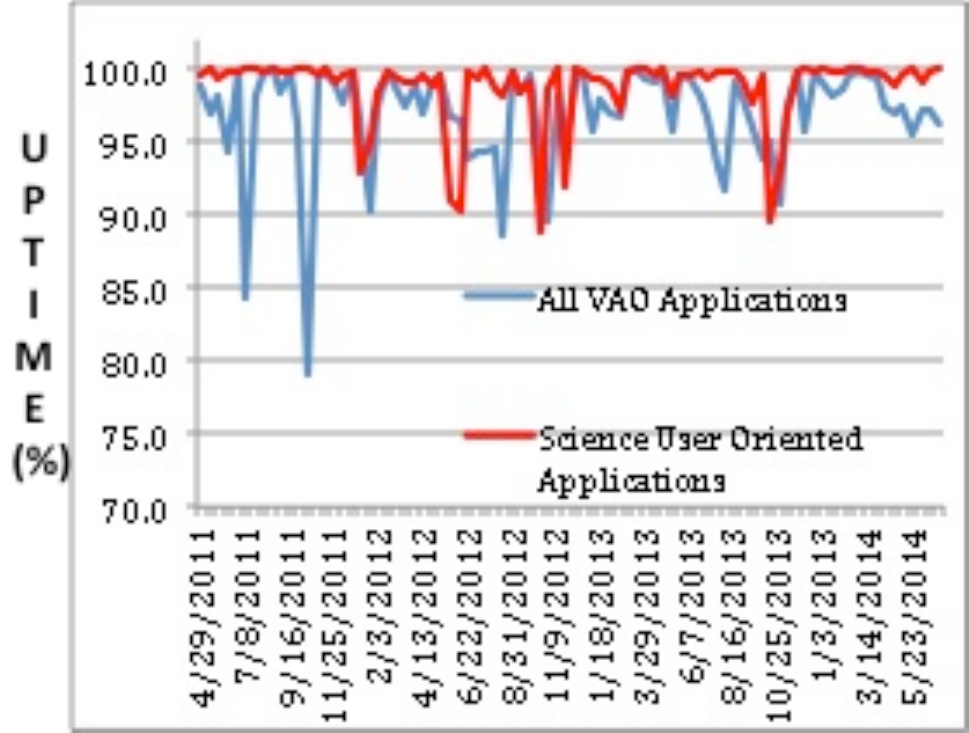}}
\caption{Operational status for all VAO services since from spring
2011 through early summer 2014 in each biweekly period.}
\label{fig:OperationsStatus}
\end{figure}

Figure~\ref{fig:OperationsStatus} shows the operational status for all
VAO services from spring 2011 through early summer 2014 in each
bi-weekly period. The blue line shows that some of the internal VAO
services---not seen directly by our science users---have had significant
downtime recently. This mostly reflects in our testing and validation
tools. More critically, the red line indicates only one significant
lapse, in~2013 October, for the science-oriented services since early 2013. This was directly due to the shutdown of US federal services that affected NASA sites.

\subsection{Monitoring and Validation of VO Data Providers}

Since the effective operation of the VAO from the perspective of
science users required that VAO data providers' services were
available, in addition to testing the aliveness of VAO services, the
VAO also monitored whether data services external to the VAO were working. Every
site that published data through the VO was tested each hour. Not all
published services were tested; rather a representative service from
each of class of services at a site was tested. All tests were
recorded and the current status of all VO sites could be seen at the
VO monitoring web site. When a problem was detected, the VAO
operations monitor contacted the responsible party and noted the
problem. In many cases the VAO assisted such sites in rapidly bringing their services back on-line.

Occasionally a VO data-providing site is abandoned. When sites were not
responsive after two months, the VAO monitoring service
deprecated them in the VAO registry so that users would no longer see them in typical queries.

Each week approximately 5--10 service interruption issues were
handled. In addition to testing whether services were available, the
VAO also validated every published VO service using the
catalog/table, image, spectral, or registry service validators. 
Each day approximately 300 services were
validated and all validation issues were recorded in a database. This
means that all published services were validated roughly once per
month. Periodically, a summary report describing the VAO validation
issues was prepared for each site, in order to provide concrete recommendations for resolution of validation issues.

A service that does not pass full validation can still provide valuable information, but obtaining more complete agreement with the IVOA standard ensures that tools work more robustly.

\begin{figure}[tb]
\centering
\resizebox{\hsize}{!}{\includegraphics{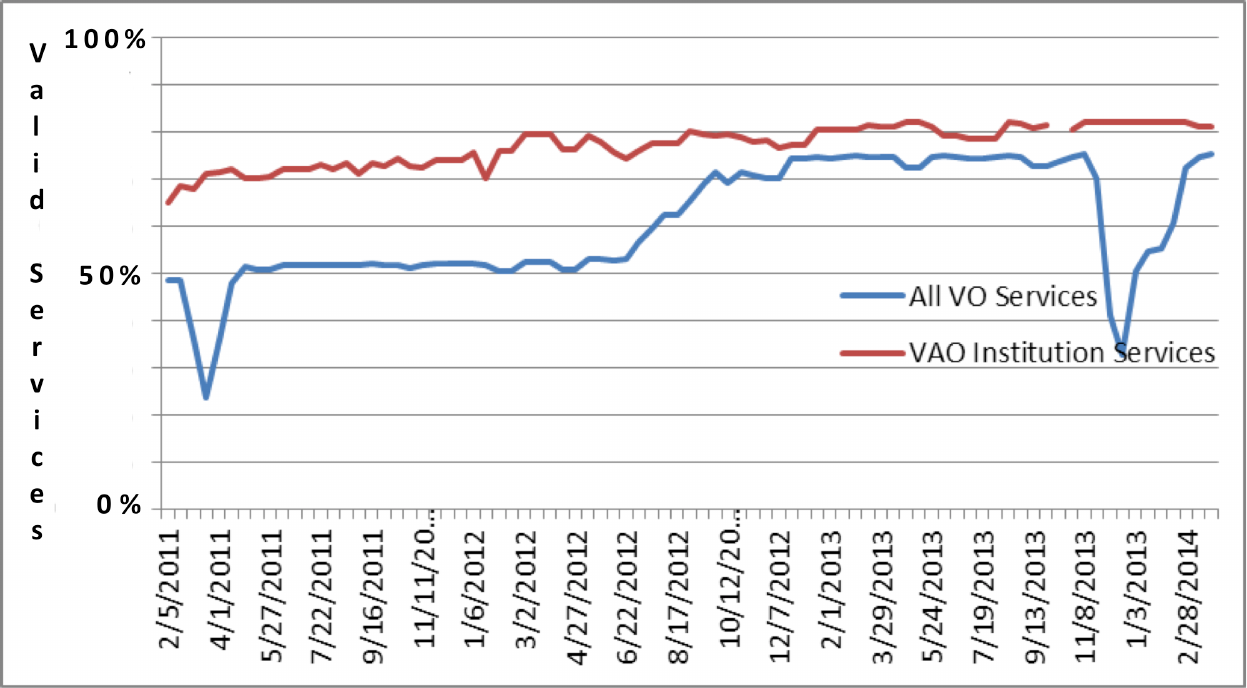}}
\caption{Variation with time of the fraction of VO services that
completely passed validation.}
\label{fig:ValidationStatus}
\end{figure}

Figure~\ref{fig:ValidationStatus} shows the fraction of VO services
that completely passed validation. The blue line shows all VO data
providers, while the red line shows the services associated with
institutions that were part of the VAO.
In
both cases there was a steady rise in compliance over the past several years. Two major drops in the overall compliance reflect bugs introduced at one of the major VO data providers outside the \hbox{VAO}. Seeing these declines our operations monitor worked with the provider, identifying specific services that were affected after initial bug fixes did not completely rectify the problem, and helped in their recovery.

\subsection{Post-VAO Operations}

The disposition of VAO-developed services and resources is discussed in detail in other sections of this paper.  Most science-oriented services will continue to be maintained by the existing institutions. The state of the internal VO services, mailing lists, documentation, blogs, and such will be maintained in the software repository. Critical infrastructure services, the web site, registry, and monitoring tools will be maintained as part of a coordinated NASA follow-on effort. This will also include at least some coordination of NASA VO operations efforts, but the level of this has yet to be determined. Our experience has shown that the \hbox{VO}, a broadly distributed system, greatly benefits from clear and comprehensive mechanisms to identify and resolve operational issues. While the NASA follow-on effort may provide some minimal capabilities, it requires a broader national and international visibility. This is not currently something that is handled by the \hbox{IVOA}.

\section{Community Engagement and User Support}

During the course of the \hbox{VAO}, effort was undertaken to ensure that products and services delivered were robust and usable by research scientists and to reach out to the broader astronomical community. The outreach efforts aimed to expose VO products and services to potential users, to assist in the take-up of those products and services, and to gather feedback in order to assure the maximum utility of the VO for astronomical research. This section describes the full scope of the efforts.

\subsection{Web Site}
\begin{figure}[tb]
\centering
\resizebox{\hsize}{!}{\includegraphics{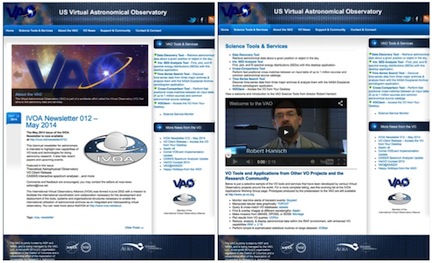}} 
\caption{(\textit{Left}) VAO web site home page.
(\textit{Right}) Science Tools \& Services area within the VAO web site. Tools and services developed by the VAO appear at the top of the document, VO tools and services provided by the community appear as well.}
\label{fig:HomePage}
\end{figure}

Figure~\ref{fig:HomePage} shows the VAO web site, with an intended audience of professional astronomers and software developers. The web site was designed both to serve as an entry portal to the VAO and to provide a means for astronomers to find information about the VO---of the more than 3 million results of a search for ``virtual observatory'' with Google, the VAO web site is one of the top hits.

From the perspective of the end user, the web site had two key areas. The first was ``Science Tools \& Services.''  This web document provided access to the web services or software developed by the \hbox{VAO}. Further, as the project began to mature, community provided tools or services began to be developed, and links to those tool or services were added.

\begin{figure}[tb]
\centering
\resizebox{\hsize}{!}{\includegraphics{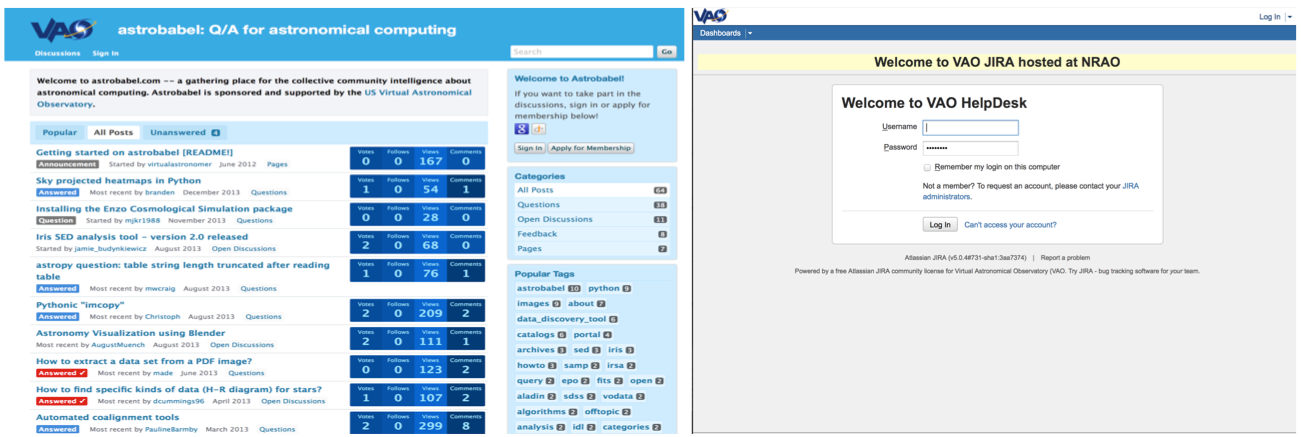}} 
\caption{(Left) VAO Forum at astrobabel.com, where users could post questions and interact with other users. (Right)  VAO Help Desk. }
\label{fig:UserSupport}
\end{figure}

The second area of interest for end users was ``Support \& Community.''  Analogous to the ``knowledge base'' that might be provided by a commercial software provider, this area was designed to help users find answers to their questions, contact other users, or submit bug reports (Figure~\ref{fig:UserSupport}).

\subsection{Product Testing}

At the beginning of the \hbox{VAO}, quality control and testing activities were under the purview of User Support. The motivation for this structure was that User Support could serve as a proxy for the end user and ensure that the products and services could be used in a research setting. For most testing activities, the User Support role was to act as the coordinator of the activities and as reviewers. In addition, User Support took the lead for performing User Acceptance Testing (UAT), which was used, along with other tests and quality control reports, to prepare software release readiness reviews.

\subsection{Documentation}

User Support staff wrote or completed user documentation in order to help research scientists have a better understanding of VAO services and applications and how to use them. Documentation packages included deployment instructions, general descriptions, tutorials, cookbooks, and similar documents. The User Support staff and product developers also collaborated to produce video tutorials, which were then made available through a YouTube channel. All software documentation produced is available in the VAO Repository  and the video tutorials remain available through YouTube.\footnote{
\texttt{http://www.youtube.com/user/usvaoTV}}

\subsection{Scientific Collaborations}

During the course of the project, the VAO supported the scientific or technical work of multiple individuals or collaborations. 
{%
The objectives of explicitly supporting such scientific collaborations
was two-fold.  First, we aimed to provide examples of the VO
infrastructure and capabilities being used for astronomical research.
Second, the interactions with the teams were anticipated to provide
feedback to the development teams for improvements to the VO
infrastructure and tools.}
The requests for support resulted both from \textit{ad hoc} proposals to the VAO and from a formal call for proposals that the VAO issued in~2012. The following is a summary of the projects and work supported.

\begin{itemize}
\item ``Real-Time Analysis of Radio Continuum Images and Time Series for ASKAP'' (PI: T.~Murphy).  This proposal requested assistance in describing multi-dimensional radio wavelength data and publishing it to the \hbox{VO}. Interaction with this team was used as a key use case in developing the VAO Standards \& Infrastructure effort toward multi-dimensional data and in interactions with the \hbox{IVOA}.
\item ``Integration of AAVSO Data Archives into the Virtual Astronomical Observatory'' (PI: M.~Templeton).  This proposal requested assistance in publishing data from the American Association of Variable Star Observers into the \hbox{VO}. The VAO provided assistance to the \hbox{AAVSO}, and the data are now available.
\item ``Cosmic Assembly/Near-infrared Deep Extragalactic Legacy Survey (CANDELS)'' (PIs: S.~Faber and H.~Ferguson).  The VAO supported the CANDELS program by distributing supernovae detections with the VOEvent network and providing access to CANDELS images through standard VO image access protocols. CANDELS supported the VAO program by providing guidance on requirements for SED building and analysis tools.
\item ``Brown Dwarf Candidate Identification Through Cross-Matching''
(PI: S.~Metchev).  The VAO supported a project that continued a search
for extremely red L- and T-type brown dwarfs that had begun during the
\hbox{NVO}. It involved cross-comparing the 2MASS and SDSS catalogs to
identify candidates that were followed-up with spectroscopy at the
Infrared Telescope Facility, Mauna Kea. The project identified the two
reddest known L~dwarfs, nine probable binaries, six of which were new
and eight of which likely harbor T~dwarf secondary stars, and  derived
an estimate of the space density of T~dwarfs \citep{2011ApJ...732...56G}.
\end{itemize}

In addition to these scientific collaborations, a scientifically motivated sub-award was issued to produce a cross-matched multi-wavelength catalog of more than 1M objects within a $10^{\circ}$ radius of the SMC was produced (``A Catalog of Spectral Energy Distributions of Stars in the Small Magellanic Cloud,'' PI: B.~Madore). The catalog is in the VAO Repository, and it has been incorporated into NED with value-added content.

\subsection{Booths and Exhibits at American Astronomical Society Meetings}\label{sec:aas}

\begin{figure}[tb]
\centering
\resizebox{\hsize}{!}{\includegraphics{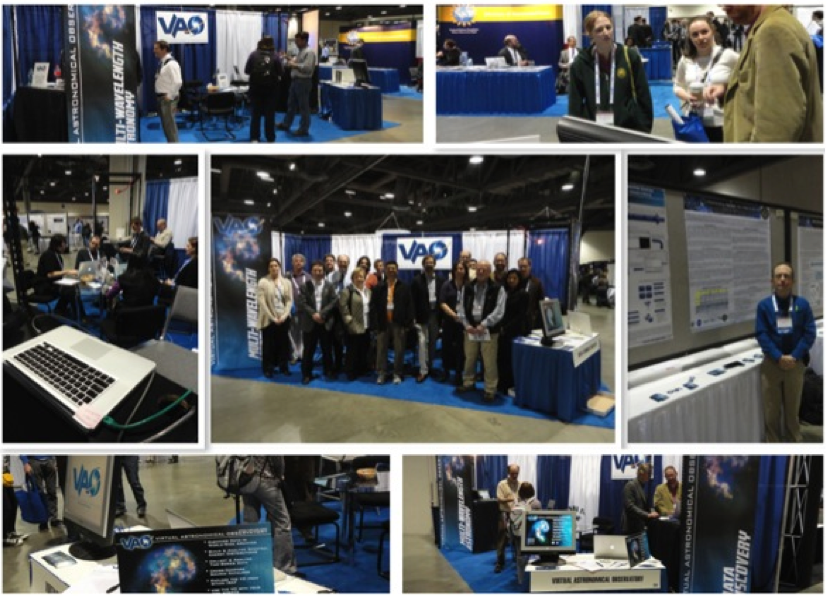}} 
\caption{Collage of images from the VAO Booth at the $221^{\mathrm{st}}$ American Astronomical Society Meeting, Long Beach, CA (2013 January). Also shown is one of the VAO-related posters \citep{2013AAS...22124037K}.}
\label{fig:AASExhibit}
\end{figure}

American Astronomical Society (AAS) meetings, principally those occurring during the winter, are one of the focal points for the U.{}S.\ (and international) astronomical community. During the course of the project, the VAO had exhibit booths at AAS meetings (Figure~\ref{fig:AASExhibit}). The use of an exhibit booth built on experience gained from NASA Archives and National observatories, for which it was found that substantial fractions of the community could be engaged at low cost. As an illustration of the value of an AAS meeting, people stopping at the exhibit were offered the opportunity to sign up for the VAO mailing list. At each AAS meeting, the size of the VAO mailing list increased by approximately 20\%.

\subsection{VAO Community Days}

\begin{figure}[t]
\centering
\resizebox{\hsize}{!}{\includegraphics{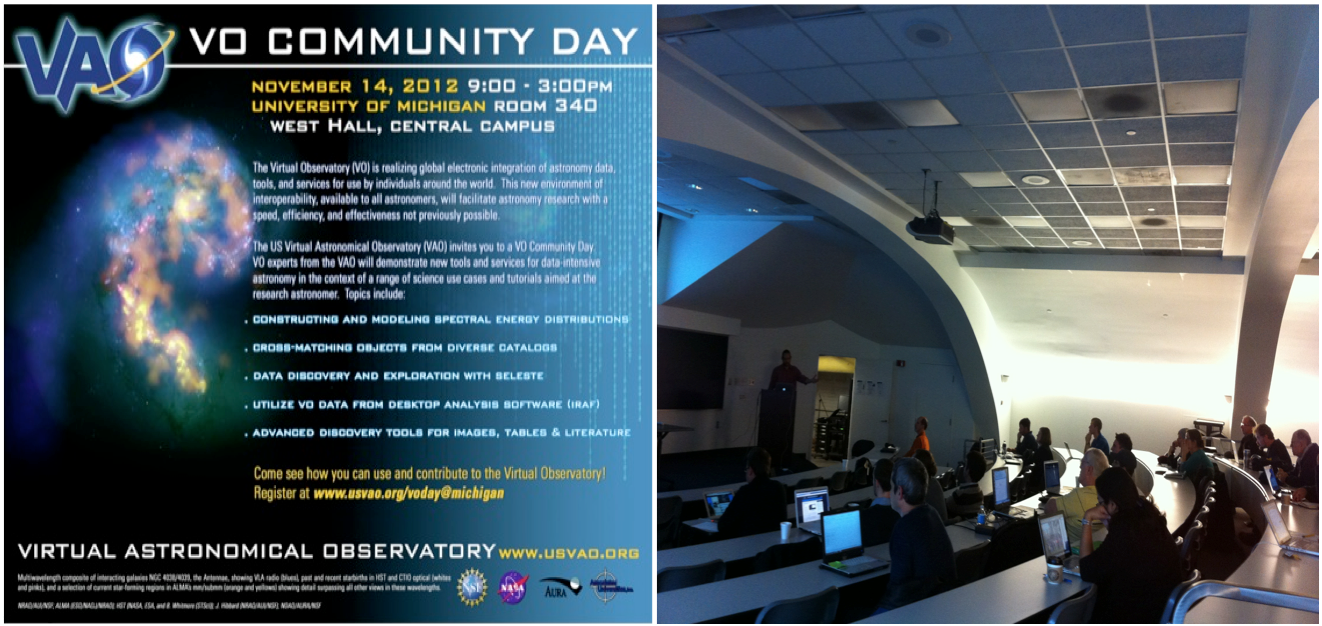}} 
\caption{(Left)  Example of an announcement flyer for a VO Community Day. (Right) Scene from a VAO Community Day, in which a VAO team member is demonstrating how a VO tool could be used. Both of these examples are from the Community Day held at the University of Michigan.}
\label{fig:VODays}
\end{figure}

VAO Community Days were a series of presentations and hands-on
activities designed to take the VAO to the community, demonstrate
capabilities, develop and encourage new users, and obtain feedback on
VO tools and services (Figure~\ref{fig:VODays}). Community Days were
typically structured with a morning session led by VAO team members,
with the option of an afternoon session for attendees to ask more
detailed questions to VAO team members or to bring in their research
questions to assess how VO tools and services could assist
them. Community Days were aimed initially at locations where there
were a large number of astronomers with the goal of making it easy for
many to attend. Table~\ref{tab:VODays} lists the VO Community Days
that were held.  Two VAO Community Days (at the University of
Washington and Cornell University) were being planned when the VAO was
directed to discontinue them in preparation for its close-out activities.

\begin{table}[h]  
\caption{VAO Community Day Locations and Dates\label{tab:VODays}}
\begin{tabular}{ll} 
Center for Astrophysics, Cambridge, MA  & 2011-11-30 \\
Caltech, Pasadena, CA & 2011-12-09 \\
U.~Arizona, Tucson, AZ & 2012-03-13 \\
U.~Michigan, Ann Arbor, MI & 2012-11-14 \\
STScI, Baltimore, MD & 2012-11-27 \\ 
\end{tabular}
\end{table}

In addition to the VAO Community Days organized by the \hbox{VAO}, VAO
Team Members also participated in similar activities organized by
international organizations, {including in Italy, Brazil, and Chile}.

\subsection{Summer Schools}\label{sec:schools}

During the \hbox{VAO}, VAO Team Members participated in summer schools
organized by other institutions, often presenting lectures or
developing demonstrations.  The NVO project hosted four Summer Schools between~2004
and~2008. During these week-long intensive sessions, over~160
participants worked with experienced VO users and software specialists
to become familiar with how to discover, access, visualize, and analyze
data, and how to use the data publication and high performance
computing capabilities of the \hbox{VO}. Those attending were
introduced to VO tools and utilities and use them to accomplish a
variety of research goals including data mining, multiwavelength research, and
time domain astronomy. In the second half of the session small teams
created their own VO-enabled data analysis applications. Students were
asked to work on team-based projects using VO protocols and software
in service of astronomical science. At the end of the school when the
projects were presented, Summer School faculty granted awards to the
five best projects. Winning projects received financial support to
attend and present their work at forthcoming winter AAS meetings.

One NVO Summer School led to the production of the book \textit{The
National Virtual Observatory: Tools and Techniques for Astronomical
Research} \citep{2007ASPC..382.....G}, which contained the lectures
and tutorials from that school.  
The volume also included a complete set of software libraries and worked examples to guide the astronomer/software developer through the process of developing VO-enabled programs in a variety of programming languages and scripting environments. Several chapters describe research results obtained by participants in the NVO Summer Schools using VO tools and technologies.

\section{Long-Term Curation of VAO Assets}

\subsection{The VAO Repository}

The VAO is making available all its digital assets---including code, documentation, data-bases, reports---through a single Google Services repository, chosen because it is free of charge, stable, and openly accessible.\footnote{
\texttt{https://sites.google.com/site/usvirtualobservatory/}}
Its existence has been announced through venues such as the AAS Newsletter, the IVOA Newsletter, and astronomy blogs and social forums. The code repository will contain all builds of the VAO software components, and all the information needed to build and use them. This content includes build instructions, release history, system requirements, license information, test results, documentation, user guides, and tutorials. The material has a common organization and look-and-feel. Currently, the repository contains builds of the science application codes, the VAO single sign-on and login codes, and the monitoring and validation software. In addition, the repository mirrors snapshots of all software that has been committed to the VAO SVN development repository, via automated weekly up-dates.

The VAO chose not to have a software licensing policy as there will be no organization to enforce it after close-out. The software is therefore released as public domain software, while duly honoring institutional licensing policies and licensing restrictions implied by the licensing of dependent third-party software. Thus, the Iris SED builder developed at SAO is released with an Apache~2.0 license and the cross-comparison code developed at Caltech/IPAC is released with a BSD 3-clause license.

All completed documentation has been posted to the 
repository,\footnote{
\texttt{https://sites.google.com/site/usvirtualobservatory/home/ documents}}
including software documentation, project reports, and outreach material. All project presentations and papers are also 
available.\footnote{
\texttt{https://sites.google.com/site/usvirtualobservatory/home/ documents/publications-presentations}}
The VAO YouTube channel, blog, Facebook page and Twitter feed will remain live.

\subsection{Transition of the VAO Infrastructure to the NASA Archives}

In response to a Call for Proposals issued by NASA in~2013 August, the
NASA archives at STScI (MAST), IPAC (\hbox{NED}, \hbox{IRSA}, NASA
Exoplanet Archives) and HEASARC submitted a proposal to sustain the
core infrastructure components of the VAO within their ``in-guide''
budgets, beginning FY~2015 (2014 October~1).
That proposal was accepted, and the NASA Archives began their
activities to sustain the core VO infrastructure elements.
A Project Scientist at HEASARC will coordinate VO activities between archives and report to NASA on VO-related activities.

\section{The VAO Legacy}

The impact of the U.{}S.\ VO programs on the international VO can be seen in a number of ways:
\begin{itemize}
\item Significant contributions to at least 35 IVOA standards and documents, from the first basic standards and services (VOTable, Simple Cone Search) to sophisticated data models and advanced data access protocols (Table Access Protocol, ObsCore, SIAP Version~2, \ldots). 
\item Leadership of numerous IVOA Working Groups and Interest Groups, as well as leadership at the IVOA Executive level. 
\item A rich infrastructure for data discovery and access, with wide deployment and implementation at major data centers in the US.
\item A robust operational environment in which distributed services are routinely validated against IVOA standards.
\item A system of resource registries that enables discover of data and data services through the world.
\item Exemplar science applications for data discovery, spectral energy distribution construction and analysis, and catalog cross-comparison.
\item Desktop scripting tools including a native Python implementation.
\item Cloud-based data storage for collaborative research and simple data sharing with the research community.
\item Creation of a ``data scientist'' position at the American Astronomical Society whose responsibilities include ``to help process and manage the increasing volume of digital data and to integrate it within the Virtual Observatory.''
\item A repository of all VAO products:  software, documentation,
tutorials, videos, news-letters, \ldots.
\item An increasing expectation that new telescopes and facilities
incorporate VO capabilities during the design of their data management
systems 
\citep[e.g.,][]{2012arXiv1201.1281M,2012SPIE.8448E..0PG,2013AAS...22124701J,2013ASPC..475..231A,2014SPIE.9149E..06S}.
\end{itemize}

However, it is more difficult to measure impact quantitatively. Since
the VAO was mostly about the deployment of software tools and
infrastructure services, it can be challenging to attribute data
accesses to the VAO as opposed to the underlying data services. Web
applications are primarily entry points to VO services; scripting
environments are needed for bulk processing. In the astronomy
community at least, and probably in many other disciplines, new
software can take many years to penetrate the community, and even
then, there is not a strong culture of software citation. For example,
we find that although some 22,000 peer-reviewed papers mention the VLA
radio telescope, only 68 formally acknowledge the use of AIPS and only
59 acknowledge use of \hbox{CASA}, the two dominant reduction and
analysis packages for radio interferometry data. Remarkably (or
perhaps not, given the situation for software citation) of over 13,000
peer-reviewed publications in astronomy and astrophysics published in
2013, only 4\% acknowledge use of the \hbox{ADS} (M.~Kurtz 2014, private communication)
and the ADS is probably the most widely-used software system in the field. Thus, counting acknowledgments to VAO or VO tools is unlikely to reflect accurately on community take-up.

On the other hand, VAO usage logs indicate close to one million VO-based data accesses per month at U.{}S.\ data providers, and with $\sim\,100$ organizations who have published some 10,000 VO-compliant data services worldwide.  VAO usage logs also show some 2,000 distinct users of VAO services in the past three months (April--June 2014). The ADS lists over 2,500 papers (about half of these peer-reviewed) citing ``virtual observatory'' in some context, and these papers are read as often and cited as often as other types of papers. Of course, without reading each and every paper one cannot be sure of the level of contamination in this sample (a paper saying ``our observatory has photometry measurements of virtually thousands of stars'' would count as a hit).
A list of  $\sim\,100$ papers that make explicit use
of VO tools and services are listed at
\texttt{http://www.usvao.org/support-community/ vo-related-publications/}.

The VO concept has been adopted in numerous other fields, particular in space science (with seven VxOÕs within NASA), plus the Virtual Solar Observatory (\hbox{NASA}, NSO), Planetary Science Virtual Observatory (Europe), and the Deep Carbon Virtual Observatory (Rensselaer Polytechnic Institute). The VO concept was recently endorsed by a panel of neuroscientists convened by the Kavli Foundation and General Electric as a means for improving access and interoperability to the vast data sets being collected in the European Brain Project and U.{}S.\ Brain Initiative. VAO and IVOA participants are now playing leading roles in the international Research Data Alliance and the newly formed U.{}S.\ National Data Services Consortium.

\subsection{Lessons Learned}

In looking back over the VAO project and its NVO predecessor, a number
of ``lessons learned'' is apparent.

\begin{itemize}
\item%
Successful infrastructure is largely invisible and
unappreciated. Developing metrics for measuring the success of
software infrastructure is a difficult question that reaches across
all scientific disciplines.  The topic was, for example, discussed in
detail at the 2015 NSF Software Infrastructure for Sustained
Innovation (SI2) Principal Investigators meeting.\footnote{
\texttt{http://cococubed.asu.edu/si2pimeeting2015/index.html}}
We urge scientists and funding agencies to investigate it collectively and develop guidelines for measuring the impact of software infrastructure. More attention should have been paid to explaining the VO infrastructure to the user community and the funding agencies, and we recommend that similar projects make such explanations a priority even in the earliest phases of development.

\item%
Deployment of a distributed infrastructure takes considerable time. Community consensus and buy-in require early and ongoing participation.  The VAO team inherited the  solutions and approach of the its technology-driven predecessor,  the \hbox{NVO}, primarily because the both projects had a common core staff.  Consequently, the VAO was slow to engage the user community and deliver services that have value to astronomers in their day-to-day work. The approach eventually used, of bringing the VAO to astronomers through integration into widely used tools in consultation with the community, led to the successful delivery and take-up of the  PyVO and VOClient toolkits. Nevertheless, and an earlier start would have led to more advanced and richly-featured services.

\item%
It is important to do marketing to the research/user community, and to manage expectations. Promising too much is as bad or worse than delivering too little.  Early promises for VO capabilities were overly ambitious and led to significant skepticism.  This ambitious program was also the primary reason why the VAO was in the position of making a substantial number of deliveries  in the final three months of the project, precisely when staff are moving on new projects at their home institutions.
As a result, some deliveries were snapshots of the code rather the full featured and well documented deliveries. Thus, in addition to managing expectations, we recommend scheduling the majority of deliveries in the earlier phases of a project.  Placing all the software in a central public repository ensures that that all the code, whether full deliveries or snapshots, is available to the community for further development.

\item%
The absence of a dedicated test team, led to by a dedicated test engineer, that would be available to support development and execution of test plans across the VAO increased the overhead on managing and organizing testing. This overhead arose because test teams were assembled on-the-fly from available staff, and test plans were consequently begun late in the development phase. We recommend establishing an independent test team at the start of a project, who coordinate with developers throughout the development lifecycle.

\item%
An essential element of the VO is the Registry, within which
data providers indicate what services they provide.  Initially, an
approach of having an easy registration process was adopted, with the
consequence that some of the services registered were either of low
quality or poorly maintained.  It is difficult to achieve the correct
balance between easy registration to encourage a substantial Registry
and substantial initial quality control that results in a Registry not
containing expected services.

\item%
A distributed project has both advantages and disadvantages:
\begin{description}
\item[Advantage] Access to a diversity of skills and different environments for validating technical approaches and implementations.
\item[Disadvantage] Coordination of efforts takes time; staff members have competing priorities as most were 
not working on VAO full-time.
\end{description}
For VAO the advantages outweighed the disadvantages, though there were certainly inefficiencies resulting from the distributed nature of the development work. These inefficiencies were minimized by having staff at only two or three organizations responsible for deliveries. For example,  staff from \hbox{SAO}, \hbox{STScI}, and IPAC/NED developed Iris. Where appropriate, one organization was responsible for a component. HEASARC managed the operational monitoring system, for example, and NOAO managed the User Support system.

\item%
Setting up an independent management entity such as the
\hbox{VAO}, \hbox{LLC}, is a non-trivial effort, though in the VAO
case it proved to be worthwhile and effective.  Having a dedicated Board of Directors to provide focused advice was a great asset.

\item%
Top-down imposition of standards is likely to fail.  Attempts to turn the OpenSkyQuery protocol into an IVOA standard, for example, did not succeed because one group proposed the standard and suggested that everyone else just adopt it.

\item%
Coordination at the international level is essential, but takes time and effort.  It can be difficult to reach consensus, or even know if consensus has been reached, owing to different cultures and communications styles.

\item%
Explicit definition of data models is important, even in cases where they seem obvious.  Constructing data models after-the-fact leads to having to redefine protocols.

\item%
Metadata collection and curation are essential and ongoing tasks, but complex, and represent a considerable investment.  Across the entire \hbox{VO}, resources were never adequate to do a proper job of curation.
\end{itemize}

\section{Conclusions}

The NVO and \hbox{VAO}, working with international partners, have established the key infrastructure for data discovery, access, and interoperability in astronomy and this infrastructure extends world-wide by virtue of collaboration with the \hbox{IVOA}. This infrastructure is both widely adopted and heavily used, although because of the nature of infrastructure people are often unaware that they are using the \hbox{VO}.
The IVOA has also developed a rich body of standards---45 in all---in the remarkably short period of~12~years, and the international VO efforts remain strong. Through the transfer of VAO assets to \hbox{NASA}, with open source software and documentation, the VAO legacy will be preserved and, we hope, enhanced. The VAO legacy will also be protected through the establishment of the U.{}S.\ Virtual Observatory Alliance under the \hbox{AAS}.

\section*{Acknowledgments}

The VAO program would not have been possible without the financial
support of the National Science Foundation (AST-0834235) and
\hbox{NASA} (NNX13AC07G to STScI/MAST), and it was supported by
\hbox{NASA/HEASARC}.  Funding at IPAC has been provided by a grant
from the National Aeronautics \& Space Administration (NASA) to the Jet Propulsion Laboratory, operated by the California Institute of Technology under contract to \hbox{NASA}.
We appreciate the wise guidance of the Board of Directors of the \hbox{VAO}, \hbox{LLC}, and the VAO Science Council, and we are grateful for feedback from the astronomical community that helped us improve our science tools and infrastructure.
Part of this research was carried
out at the Jet Propulsion Laboratory, California Institute of
Technology, under a contract with the National Aeronautics \& Space Administration.

Foremost, we acknowledge the dedication, commitment, and
excellence of the VAO project team. We brought together the best of
the best, from nine different organizations, and through pursuit of
common goals created a data management infrastructure that has brought
about a sea change in how we manage and share data in astronomy and
that has become a model for data management in many other
disciplines.  We express our gratitude D.~De~Young (deceased, 2011 December) for his astute guidance
throughout the NVO Project and in the initial phases of the \hbox{VAO}.

This research has made use of NASA's Astrophysics Data System Bibliographic Services.

\appendix

\section{VAO Institutions}\label{app:institute}

The VAO was operated as a limited liability company, funded by the
National Science Foundation with coordinated funding provided by the
National Aeronautics \& Space Administration.
Table~\ref{tab:institute} lists the institutions engaged in the
scientific and technical development work of the \hbox{VAO}; business
management was provided by the Associated Universities, Inc.\ (AUI).

\begin{table*}[tbh]
\centering
\caption{Participating VAO Institutions\label{tab:institute}}
\begin{tabular}{p{0.4\textwidth}p{0.4\textwidth}}
\noalign{\hrule\hrule}
\textbf{NSF} & \textbf{NASA} \\
\noalign{\hrule}
California Institute of Technology (Caltech) 
	& High Energy Astrophysics Science Archive Research Center
	  (HEASARC) \\
Johns Hopkins University (JHU) 
	& Infrared Processing \& Analysis Center, California Institute
	  of Technology (IPAC) \\
National Center for Supercomputing Applications (NCSA) 
	& Jet Propulsion Laboratory, California Institute of
	  Technology (JPL) \\
National Optical Astronomy Observatory (NOAO)
	& Space Telescope Science Institute (STScI) \\
National Radio Astronomy Observatory (NRAO) & \\
Smithsonian Astrophysical Observatory (SAO) & \\
\noalign{\hrule\hrule}
\end{tabular}
\parbox{0.9\textwidth}{Institutions are listed according to which
agency provided the significant funding for VAO work.}
\end{table*}

\section{IVOA Standards}\label{app:standards}

This appendix lists International Virtual Observatory Alliance
standards and recommendations for which VAO Team Members were
identified either as authors or editors.  Standards and
recommendations are listed in reverse chronological order of adoption.

\begin{itemize}

\item%
``VOTable Format Definition,'' Version~1.3, IVOA Recommendation, 20~September 2013 (F.~Ochsenbein, R.~Williams, C.~Davenhall, M.~Demleitner, D.~Durand, P.~Fernique, D.~Giaretta, R.~Hanisch, T.~McGlynn, A.~Szalay, M.~Taylor, A.~Wicenec)

\item%
``Data Access Layer Interface,'' Version~1.0, IVOA Recommendation, 29~November 2013 (P.~Dowler, M.~Demleitner, M.~Taylor, D.~Tody)

\item%
``IVOA Registry Relational Schema,'' Version~1.0, IVOA Proposed Recommendation, 27~February 2014 (M.~Demleitner, P.~Harrison, M.~Molinaro, G.~Greene, T.~Dower, M.~Perdikeas)

\item%
``MOC - HEALPix Multi-Order Coverage map,'' Version~1.0, IVOA Proposed Recommendation, 10~March 2014 (T.~Boch, T.~Donaldson, D.~Durand, P.~Fernique, W.~O'Mullane, M.~Reinecke, M.~Taylor)

\item%
``Simple Application Messaging Protocol,'' Version~1.3  IVOA Recommendation, 11~April 2012 (M.~Taylor, T.~Boch, M.~Fitzpatrick, A.~Allan, J.~Fay, L.~Paioro, J.~Taylor, D.~Tody)

\item%
``Simple Line Access Protocol,'' Version~1.0, IVOA Recommendation, 09~December 2010 (J.~Salgado, P.~Osuna, M.~Guainazzi, I.~Barbarisi, M.-L.~Dubernet, D.~Tody)

\item%
``Simple Spectral Access Protocol,'' Version~1.1, IVOA Recommendation, 10~February 2012 (D.~Tody, M.~Dolensky, J.~McDowell, F.~Bonnarel, T.~Budavari, I.~Busko, A.~Micol, P.~Osuna, J.~Salgado, P.~Skoda, R.~Thompson, F.~Valdes, and the Data Access Layer working group)

\item%
``Table Access Protocol,'' Version~1.0, IVOA Recommendation, 27~March 2010 (P.~Dowler, G.~Rixon, D.~Tody)

\item%
``TAPRegExt: a VOResource Schema Extension for Describing TAP Services,'' Version~1.0, IVOA Recommendation, 27~August 2012 (M.~Demleitner, P.~Dowler, R.~Plante, G.~Rixon, M.~Taylor)

\item%
``IVOA Spectral Data Model,'' Version~2.0, IVOA Proposed Recommendation, 09~March 2014 (J.~McDowell, D.~Tody, T.~Budavari, M.~Dolensky, I.~Kamp, K.~McCusker, P.~Protopapas, A.~Rots, R.~Thompson, F.~Valdes, P.~Skoda, B.~Rino, S.~Derriere, J.~Salgado, O.~Laurino, and the IVOA Data Access Layer and Data Model Working Groups)

\item%
``Observation Data Model Core Components and its Implementation in the Table Access Protocol,'' Version~1.0, IVOA Recommendation, 28~October 2011 (M.~Louys, F.~Bonnarel, D.~Schade, P.~Dowler, A.~Micol, D.~Durand, D.~Tody, L.~Michel, J.~Salgado, I.~Chilingarian, B.~Rino, J.~de~Dios Santander, P.~Skoda)

\item%
``VOSpace Specification,'' Version~2.0, IVOA Recommendation, 29~March 2013 (M.~Graham, D.~Morris, G.~Rixon, P.~Dowler, A.~Schaaff, D.~Tody)

\item%
``IVOA Credential Delegation Protocol,'' Version~1.0, IVOA Recommendation, 18 February 2010 (M.~Graham, R.~Plante, G.~Rixon, G.~Taffoni)

\item%
``Web Services Basic Profile,'' Version~1.0, IVOA Recommendation, 16~December 2010 (A.~Schaaff, M.~Graham)

\item%
``StandardsRegExt: a VOResource Schema Extension for Describing IVOA
Standards,'' Version~1.0, IVOA Recommendation, 08 May 2012 (P.~Harrison, D.~Burke, R.~Plante, G.~Rixon, D.~Morris, and the IVOA Registry Working Group)

\item%
``Describing Simple Data Access Services,'' Version~1.0, IVOA Recommendation, 25~November 2013 (R.~Plante, J.~Delago, P.~Harrison, D.~Tody, and the IVOA Registry Working Group)

\item%
``VODataService: a VOResource Schema Extension for Describing Collections and Services,'' Version~1.1, IVOA Recommendation, 02~December 2010 (R.~Plante, A.~St{\'e}b{\'e}, K.~Benson, P.~Dowler, M.~Graham, G.~Greene, P.~Harrison, G.~Lemson, T.~Linde, G.~Rixon)

\item%
``IVOA Registry Relational Schema,'' Version~1.0, IVOA Proposed Recommendation, 27~February 2014 (M.~Demleitner, P.~Harrison, M.~Molinaro, G.~Greene, T.~Dower, M.~Perdikeas)

\item%
``IVOA Document Standards,'' Version~1.2, IVOA Recommendation, 13 April 2010 (R.J.~Hanisch, C.~Arviset, F.~Genova, B.~Rino)

\item%
``Sky Event Reporting Metadata,'' Version~2.0, IVOA Recommendation, 11~July 2011 (R.~Seaman, R.~Williams, A.~Allan, S.~Barthelmy, J.~Bloom, J.~Brewer, R.~Denny, M.~Fitzpatrick, M.~Graham, N.~Gray, F.~Hessman, S.~Marka, A.~Rots, T.~Vestrand, P.~Wozniak)

\item%
``IVOA Support Interfaces,'' Version~1.0, IVOA Recommendation, 31~May 2011 (Grid and Web Services Working Group, M.~Graham, G.~Rixon)
\end{itemize}

\clearpage

\section{International Virtual Observatory Alliance
	Leadership}\label{app:ivoa_lead}

Table~\ref{tab:ivoa.exec} lists VAO Team members who served in various leadership
positions within the \hbox{IVOA}.

\begin{table*}[t]
\centering
\caption{VAO Leadership within the IVOA\label{tab:ivoa.exec}}
\begin{tabular}{llll}
\noalign{\hrule\hrule}
 & \textbf{Position} & \textbf{Individual} & \textbf{Term} \\
\noalign{\hrule}
\multirow{4}{0.2\textwidth}{Executive Committee}
	& Chair        & R.~Hanisch  & 2002 June--2003 July \\
	& Deputy Chair & D.~De~Young & 2006 August--2007 August \\
	& Chair        & D.~De~Young & 2007 August--2008 October \\
	& Secretary    & J.~Evans    & 2013 September--2014 September \\
\noalign{\hrule}

{Technical Working Group}
	& Chair & R.~Williams & 2002 June--2006 July \\

\noalign{\hrule}

\multirow{2}{0.2\textwidth}{Technical Coordination Group}
	& Chair        & R.~Williams & 2006 July--2008 May \\
	& Deputy Chair & M.~Graham   & 2012 May--2014 September \\

\noalign{\hrule}

\multirow{3}{0.2\textwidth}{Inter-operability Conference Program Organizing Committee}
	& Member & R.~Hanisch & 2003 March--2007 May \\
	& Member & M.~Graham  & 2012 May--2014 September \\
	&        &            & \\

\noalign{\hrule}

{Standards and Process Subcommittee}
	& Member  & R.~Hanisch & 2007 September--2010 September \\

\noalign{\hrule}

{Document Coordinator}
	& & S.~Emery Bunn & 2010 July--2014 September \\

\noalign{\hrule}

\multirow{2}{0.2\textwidth}{Applications Working Group}
	& Chair      & Tom McGlynn   & 2008 July--2011 July \\
	& Vice Chair & Tom Donaldson & 2014 May--2014 September \\

\noalign{\hrule}

\multirow{2}{0.2\textwidth}{Data Access Layer Working Group}
	& Chair      & Doug Tody        & 2003 June--2007 May \\
	& Vice Chair & Mike Fitzpatrick & 2010 May--2013 May \\

\noalign{\hrule}

\multirow{3}{0.2\textwidth}{Data Models Working Group}
	& Chair      & Jonathan McDowell & 2003 June \\
	& Vice Chair & Omar Laurino      & 2011 May--2014 May \\
	& Vice Chair & Omar Laurino      & 2014 May--2014 September \\

\noalign{\hrule}

\multirow{2}{0.2\textwidth}{Grid \& Web Services Working Group}
	& Chair & Matthew Graham & 2006 December--2007 May \\
	& Chair & Matthew Graham & 2007 May--2011 May \\

\noalign{\hrule}

\multirow{4}{0.2\textwidth}{Registry Working Group}
	& Chair      & Ray Plante      & 2006 September--2009 September \\
	& Chair      & Ray Plante      & 2009 November--2010 November \\
	& Vice Chair & Gretchen Greene & 2009 November--2010 November \\
	& Chair      & Gretchen Greene & 2011 January--2014 May \\

\noalign{\hrule}

{Standards \& Processes Working Group}
	& Chair & Bob Hanisch & 2003 June--2006 May \\

\noalign{\hrule}

{Uniform Content Descriptors Working Group}
	& Chair & Roy Williams & 2003 June--2005 January \\

\noalign{\hrule}

\multirow{5}{0.2\textwidth}{VO Event Working Group}
	& Chair      & Roy Williams   & 2005 January--2008 January \\
	& Chair      & Rob Seaman     & 2006 December--2008 May \\
	& Chair      & Rob Seaman     & 2008 May--2011 May \\
	& Vice Chair & Roy Williams   & 2010 October--2011 October \\
	& Chair      & Matthew Graham & 2011 October--2012 October \\

\noalign{\hrule}

{Applications Interest Group}
	& Chair & Tom McGlynn & 2004 January--2005 July \\
	
\noalign{\hrule}

\multirow{3}{0.2\textwidth}{Data Curation \& Preservation Interest Group}
	& Chair & Bob Hanisch       & 2007 May--2010 May \\
	& Chair & Alberto Accomazzi & 2010 May--2014 May \\
	&        &            & \\

\noalign{\hrule}

{Knowledge Discovery in Databases Interest Group}
	& Chair & George Djorgovski & 2012 October--2014 September \\

\noalign{\hrule}

\multirow{2}{0.2\textwidth}{Time Domain Interest Group}
	& Chair      & Matthew Graham   & 2012 October--2013 May \\
	& Vice Chair & Mike Fitzpatrick & 2013 May--2014 September \\

\noalign{\hrule}
\noalign{\hrule}
\end{tabular}

\parbox{0.9\textwidth}{The term of the Chair of the Executive Committee was increased to~18~months
beginning in 2007 August.}
\parbox{0.9\textwidth}{In~2005 July, the Technical Working Group was
reformulated as the Technical Coordination Group.}

\parbox{0.9\textwidth}{The Chair and Vice Chair of the Data Models Working Group were both granted one year extensions in~2014 May.}

\parbox{0.9\textwidth}{The Standards \& Processes Working Group was
deactivated in~2005 May.}

\parbox{0.9\textwidth}{The Uniform Content Descriptors Working Group
was renamed to the Semantics Working Group in~2005 October.}

\parbox{0.9\textwidth}{The VO Event Working Group was converted to
the Time Domain Interest Group in~2012 October.}

\parbox{0.9\textwidth}{The Applications Interest Group was converted
to the Applications Working Group in~2007 January.}

\end{table*}

\clearpage

\bibliographystyle{elsarticle-harv}
\bibliography{VAOforA&C}

\end{document}